\documentclass[10pt]{article}

\usepackage{amsmath}
\usepackage{amssymb}

\usepackage{graphicx}

\usepackage{cite}

\usepackage{color} 

\usepackage{setspace} 
\usepackage[labelfont=bf,labelsep=period,justification=raggedright]{caption}

\makeatletter
\renewcommand{\@biblabel}[1]{\quad#1.}
\makeatother

\date{}

\newcommand{\MI}{Mutual Information}
\newcommand{\CMI}{conditional Mutual Information}
\newcommand{\IM}{Ising model}
\newcommand{\TE}{Transfer Entropy}
\newcommand{\cau}{`causality'}
\newcommand{\cs}{complex systems}

\begin{document}

\begin{flushleft}
{\Large
\textbf{Quantifying `causality' in complex systems: Understanding Transfer Entropy}
}
\\
Fatimah Abdul Razak$^{1,2,*}$, 
Henrik Jeldtoft Jensen$^{1}$, 
\\
\bf{1} Complexity \& Networks Group and Department of Mathematics, Imperial College London, 
 SW7 2AZ London, UK
\\
\bf{2} School of Mathematical Sciences, Faculty of Science \& Technology, Universiti Kebangsaan Malaysia, 43600 UKM Bangi, Malaysia
\\
$\ast$ E-mail: fatima84@ukm.my, fatimah84@gmail.com
\end{flushleft}

\section*{Abstract}
`Causal' direction is of great importance when dealing with complex systems. Often big volumes of data in the form of time series are available and it is important to develop methods that can inform about possible causal connections between the different observables. Here we investigate the ability of the Transfer Entropy measure to identify causal relations embedded in emergent coherent correlations. We do this  by firstly applying \TE\ to an amended \IM.  In addition we use a simple Random Transition model to test the reliability of Transfer Entropy as a measure of `causal' direction in the presence of stochastic fluctuations.  In particular we systematically study the effect of the finite size of data sets.


\section*{Introduction}
Many complex systems are able to self-organise into a critical state \cite{per,kim}. In the critical state local distortions can propagates throughout the entire system  \cite{kim,h,gunnar}. This leads to correlation spanning across the entire system. We address here how to identify directed stochastic causal connections embedded in stochastic fluctuating but strongly correlated background. 

Most of  \cau\ and directionality measures have been tested on low dimension systems and neglect addressing the behaviour of systems consisting of large numbers of interdependent degrees of freedom that is a main feature of \cs. From a \cs\ point of view, on one hand there is the system as a whole (collective behaviour) and on another there  are individual interactions that lead to the collective behaviour. A measure that can  help understand and differentiate these two elements is needed. We shall first seek to make a clear definition of \cau\ and then relate this definition to \cs. We outline the different approaches and measures used to quantify this type of \cau. We highlight that for multiple reasons, \TE\ seems to be a very suitable candidate for a \cau\ measure for \cs. Consequently we seek to shed some light on the usage \TE\ on \cs.

To improve our understanding of \TE\ we study two simplistic models of complex systems which in a very controllable way generates correlated time series. Complex system whose main characteristic consist in essential  cooperative behaviour~\cite{pjE09} takes into account instances when the whole system is interdependent. Therefore, we  apply \TE\ to the (amended) \IM\ in order to investigate its behaviour at different temperatures particularly near the critical temperature. Moreover, we are also interested in investigating the different magnitude of \TE\ in general (which is not fully understood \cite{runge12b}) by looking at the effect of different transition probabilities, or activity levels. We discuss the interpretation of the different magnitudes of the \TE\ by varying transition rates in a Random Transition model.

\section*{Quantifying `Causality'}

The quantification of \cau\ was first envisioned by the mathematician Wiener \cite{wiener56} who propounded the idea that the \cau\ of a variable in relation to another can be measured by how well the variable  helps to predict the other. In other words, variable $Y$ `causes' variable $X$ if the ability to predict $X$ is improved by incorporating information about $Y$ in the prediction of $X$. The conceptualisation of \cau\ as envisioned by Wiener was formulated by Granger \cite{granger69} leading to the establishment of the Wiener-Granger framework of \cau. This is the definition of \cau\ that we shall adopt in this paper. 

In literature, references to \cau\ take many guises. The term directionality, information transfer and sometimes even independence can possibly refer to some sort of \cau \ in line with the Wiener-Granger framework. Continuing the assumption that  $Y$ causes $X$, one would expect the relationship between $X$ and $Y$ to be asymmetric and that the  information flows in a  direction from the source $Y$ to the target $X$. One can assume that this information transfer is the unique information provided by the causal variable to the affected one.  When one variable causes another variable, the affected variable (the target) will be dependent (to certain extent) on the causal variable (the source).  There must exist a certain time lag however small between the source and the target \cite{seth11,sauer10, hausman99}, this will be henceforth referred to as the causal lag~\cite{granger69}. 
One could also say the Wiener-Granger framework of prediction based  \cau\ is equivalent to looking for dependencies between the variables at a certain causal lag.

Roughly, there are two different approaches in establishing \cau\ in a system. One approach is to make a qualified guess of a model that will fit the data, called the confirmatory approach \cite{friston11}. Models of this nature are typically very field specific and rely on particular insights into the mechanism involved. A contrasting approach known as the exploratory approach,  infers `causal' direction from the data.
This approach does not rely on any preconceived idea about underlying mechanisms and let results from data shape the directed model of the system. Most of the measures within the Wiener-Granger framework falls into this category.  One can think of the different approaches as being on a spectrum from purely confirmatory to purely exploratory.

The nature of  \cs\ calls for the exploratory approach. The abundance of data emphasises this even more so. In fact \cau\ measures in the Wiener Granger framework have been increasingly utilised on data sets obtained from \cs\ such as the brain \cite{vicente11,martini11} and financial systems \cite{mars02}. Unfortunately, most of the basic testings of the effectiveness of these measures is mostly done on dynamical systems \cite{sch00,sch02,pompe11} or simple time series, without taking into account the emergence of collective behaviour and criticality. Complex systems are typically stochastic and thus different from deterministic systems  where the internal and external influences are distinctly identified. As mentioned above, here we focus on the emergence of collective behaviour in complex systems and in particular on how  the intermingling of the collective behaviour with individual (coupled) interactions complicates the identification of `causal' relationships. Identifying a measure that is able to distinguish between these different interactions will obviously help us to improve our understanding the dynamics of \cs.


\section*{\TE}

Within the Wiener-Granger framework, two of the most popular \cau\ measure are Granger Causality (G-causality) and its nonlinear analog \TE.  G-causality and \TE\ are exploratory as their measures of causality are  based on distribution of the sampled data.  The standard steps of prediction based \cau\ that underlies  these measures can be summarized as follows. Say we want to test whether variable $Y$ causes variable $X$. The first step would be to predict the current value of $X$ using the historical values of $X$. The second step is to do another prediction where the historical values of  $Y$ and $X$ are both used to predict the current value of $X$. And the last step would be to compare the former to the latter. If the second prediction is judged to be better than the first one, then one can conclude that $Y$ causes $X$. This being the main idea, we outline why \TE\ is more suitable for \cs.

 Granger causality is the most commonly used \cau\ indicator \cite{seth11}. However, in the context of the nonlinearities of a \cs\ (collective behaviour and criticality being the main example), using G-causality may not be sufficient. Moreover, this AR framework makes G-causality  less exploratory than \TE. \TE\ was defined \cite{sch00,sch02}  as a nonlinear measure to infer directionality using the Markov property. The aim was to incorporate the properties of \MI\ and the dynamics captured by transition probabilities in order to understand the concept and exchange of information. More recently, the usage of \TE\ to detect causal relationships \cite{palus07, vej08, lungarella07} and causal lags (the time between cause and effect) has been further examined \cite{wibral13,runge12b}. Thus we are especially interested in \TE\ due to  its propounded ability  to capture nonlinearities, its exploratory nature as well as its information theoretic background that provides information transfer related interpretation. Unfortunately, some of the vagueness in terms of interpretation may cause confusion in \cs. The rest of the paper is an attempt to discuss these issues in a reasonably self-contained manner. 

\subsection*{\MI\ based measures}

Define random variables $X,Y$ and $Z$ with discrete probability distributions  $p_{X}(x),x \in \mathcal{X}$, $p_{Y}(y),y \in \mathcal{Y}$ and $p_{Z}(z),z \in \mathcal{Z}$. The entropy of $X$ is defined  \cite{sh,eif} as \begin{equation}
H(X)=-\sum_{x \in \mathcal{X}}p_{X}(x) \log\:p_{X}(x)
\end{equation}
where log to the base $e$ and $0\:log\:0=0$ is used. The joint entropy of $X$ and $Y$ is  defined as
\begin{equation}\label{JE} 
H(X,Y)=-\sum_{x \in \mathcal{X}} \sum_{y \in \mathcal{Y}} p_{XY}(x,y) \log\,p_{XY}(x,y)
\end{equation}
and the conditional entropy can be written as
\begin{align}\label{CE}
H(X|Y)&=-\sum_{x \in \mathcal{X}} \sum_{y \in \mathcal{Y}} p_{XY}(x,y) \, \log\,p_{X|Y}(x|y)
\end{align}
where $p_{XY}$ is the joint distribution and $p_{X|Y}$ is the respective conditional distribution. The \MI\ \cite{eif,g04}  is defined as 
\begin{align}\label{Isums}
I(X,Y)=H(X)-H(X|Y).
\end{align}
Taking into account conditional variables, the \CMI\ \cite{eif,palus07}
is defined as $I(X,Y|Z)=H(X|Z)-H(X|Y,Z)$. A variant of \CMI\ namely the \TE\ was first defined by Schreiber in \cite{sch00}. Let $X^{\tau}$ be the variable $X$ that is shifted by $\tau$, so that the values of $X^{\tau}(t)=  X(n-\tau)$ where $X(n)$ is the value of $X$ at time step $n$ and similarly for $Y$. 
We highlight a simple form of \TE\ where conditioning is minimal such that 
\begin{align}\label{TEtau}
T_{YX}^{(\tau)}
=I(X,Y^{\tau}|X^{1})=H(X|X^{1})-H(X|X^{1},Y^{\tau}).
\end{align}
The idea is that, if $Y$ causes $X$ at causal lag $t_Y$, then $T_{YX}^{(t_Y)} \ge T_{YX}^{(\tau)} $ for any lag $\tau$ since $H(X|X^{1},Y^{t_Y}) \le H(X|X^{1},Y^{\tau})$ due to the fact that  $Y^{t_Y}$ should provide the most information about the change of $X^{1}$ to $X$.
This simple form allows us to vary the values of time lag $\tau$ in ascertaining the actual causal lag. This form of \TE\ was also used in \cite{nichols05,vicente11,wibral13,li11,pompe11}. The \TE\ in equation (\ref{TEtau}) can also be written as
\begin{align}
T_{YX}^{(\tau)}
=\sum_{x \in \mathcal{X}}\sum_{x' \in \mathcal{X}}\sum_{ y\in \mathcal{Y}}p_{X X^{1} Y^{\tau}}(x,x',y)
\,\log\,\frac{p_{X| X^{1} Y^{\tau}}(x|x',y)}
{p_{X| X^{1}}(x|x')}. \end{align}
Our choice of this simple definition was motivated by the fact that it directly captures how the state of $Y^\tau(n)=Y(n-\tau)$ influences the changes in $X$ i.e. from $X(n)$ to $X^1(n)=X(n-1)$. In other words, equation (\ref{TEtau}) is tailor made to measure whether the state of $Y(n-\tau)$ influences the current changes in $X$. This coincides with the predictive view of \cau\ in the Wiener-Granger framework where the current state of one variable (the source) influences the changes in another variable (the target) in the future. The same concept will be applied in order to probe this kind of  \cau\ in our models.

\section*{The \IM}
A system is critical when correlations are long ranged.  A simple prototype example is the \IM \ \cite{kim} at critical temperature, $T_c$, away from $T_c$ correlations are short ranged and dies off exponentially with separation.  We shall apply \TE\ to the \IM\ in order to investigate its behaviour at different temperatures particularly in the vicinity of  the critical temperature. One can visualize the   2D \IM\ as a two dimensional square lattice with length $L$ composed of $N=L^2$ sites  $s_i, i \in \mathcal{N}=\{1 \cdots N\}$. These sites can only be in two possible states, spin-up ($s_i=1$) or spin-down ($s_i=-1$).
 We restrict the interaction of the sites to only its nearest neighbours (in two dimensions this will be sites to the north, south, east and west). Let the interaction strength between $i$ and $j$ be denoted by 
\begin{align}\label{isingJ}
J_{ij}= 
\begin{cases}
J \ge 0, & \text{if $i$ and $j$ are nearest neighbours and }i,j \in\mathcal{N}  \ \\
0, & \text{otherwise}
\end{cases} 
\end{align}
 so that the Hamiltonian (energy), $\mathcal{H}$, is given by \cite{kim, cipra87}
\begin{align}\label{hamiltonian}
\mathcal{H}=-\sum_{i \in \mathcal{N}\ }\sum_{j \in \mathcal{N}} J_{ij} s_is_j.
\end{align}
$\mathcal{H}$ is used to the obtain the Boltzmann (Gibbs) distribution
$
\gamma_B=\frac{\exp(-\beta\mathcal{H})}
{\sum\exp(-\beta\mathcal{H})}
$
where $\beta=\frac{1}{K_B
T}$ and $K_B$ is the Boltzmann constant and $T$ is temperature.

We implement the usual Metropolis Monte Carlo (MMC) algorithm
\cite{kim, krauth, norris} for the simulation of the Ising model in two dimensions
with periodic boundary conditions. The algorithm proposed by Metropolis and co-workers in 1953 was designed to sample the Boltzmann distribution $\gamma_B$ by artificially imposing dynamics on the \IM. The implementation of the MMC algorithm in this paper is outlined as follows.
A site is chosen at random to be considered for flipping (change of state) with
probability $\gamma_B$. The event of considering the change and afterwards the actual change (if accepted) of the configuration, shall henceforth be referred to  as flipping consideration.
A sample is taken after each $N$ flipping considerations. The logic being that, since sites to be considered are chosen randomly one at a time,  after $N$ flips, each site will on average have been selected for consideration once.  The interaction strength is set to be $J=1$ and the Boltzmann constant is fixed as $K_B=1$ for all the simulations. We let the system run up to $2000$ samples before sampling at every $N=L^2$ time steps. 

Through the MMC algorithm, a Markov chain (process) is formed for every site on the lattice.
 The state of each site at each sample will be taken as a time step $n$ in the Markov chain $(s_X )_n$. Let $S$ be the number of samples (length of the Markov chains). To get the probability values for each site, we utilise temporal average. All the numerical probabilities obtained for the \IM\ in this paper have been obtained by averaging   over simulations with $S=100000$ unless stated otherwise.

\subsection*{Measures on \IM}\label{MIM}

\begin{figure}[ht]
\begin{minipage}[b]{0.5\linewidth}
\centering
\includegraphics[width=\textwidth]{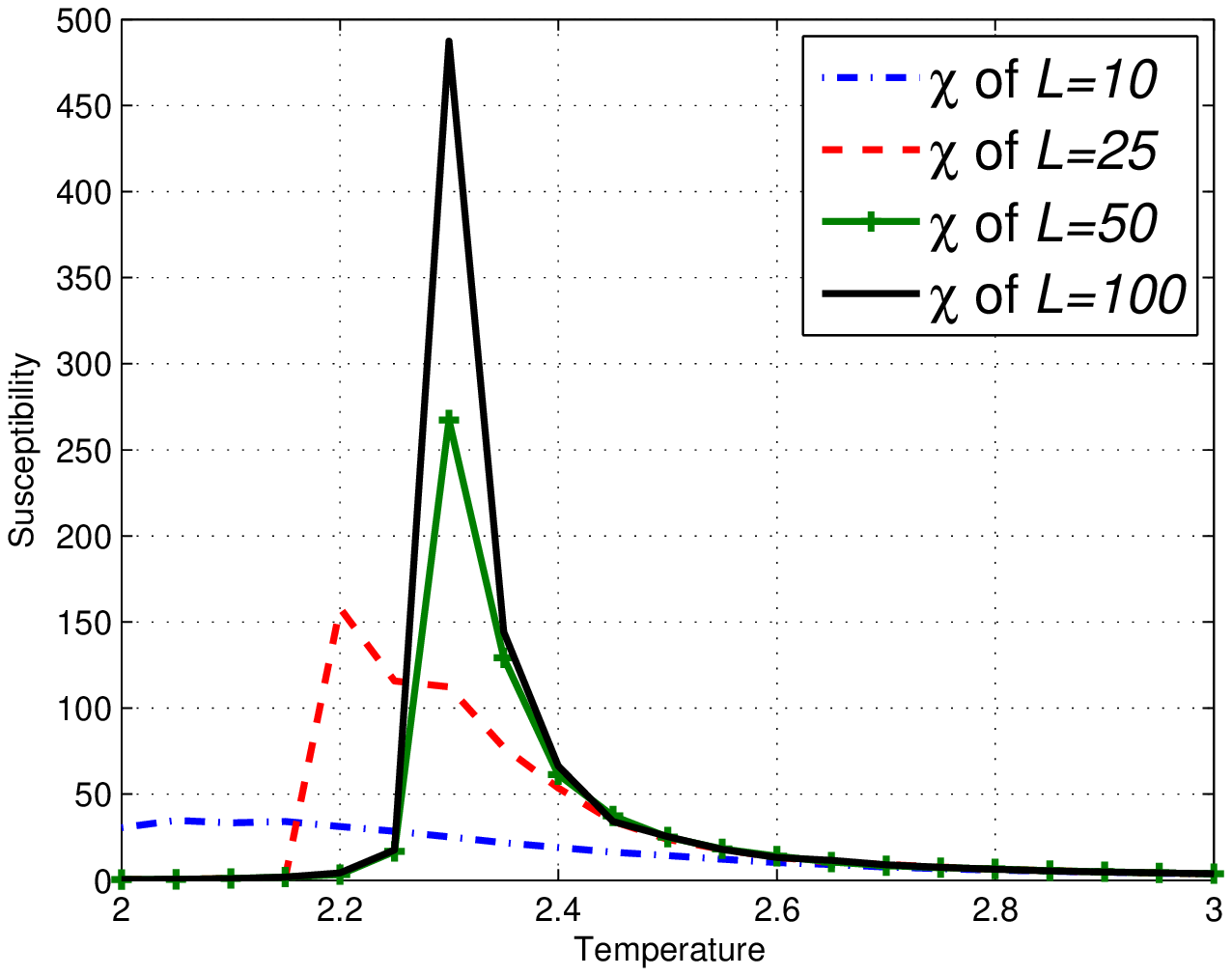}
\caption{\textbf{Susceptibility $\chi$  on the \IM} with lengths $L=10,25,50,100$ obtained using equation (\ref{susceptibility}). Peaks can be seen at respective $T_c$.}
\label{suscep100}
\end{minipage}
\hspace{0.3cm}
\begin{minipage}[b]{0.5\linewidth}
\centering
\includegraphics[width=\textwidth]{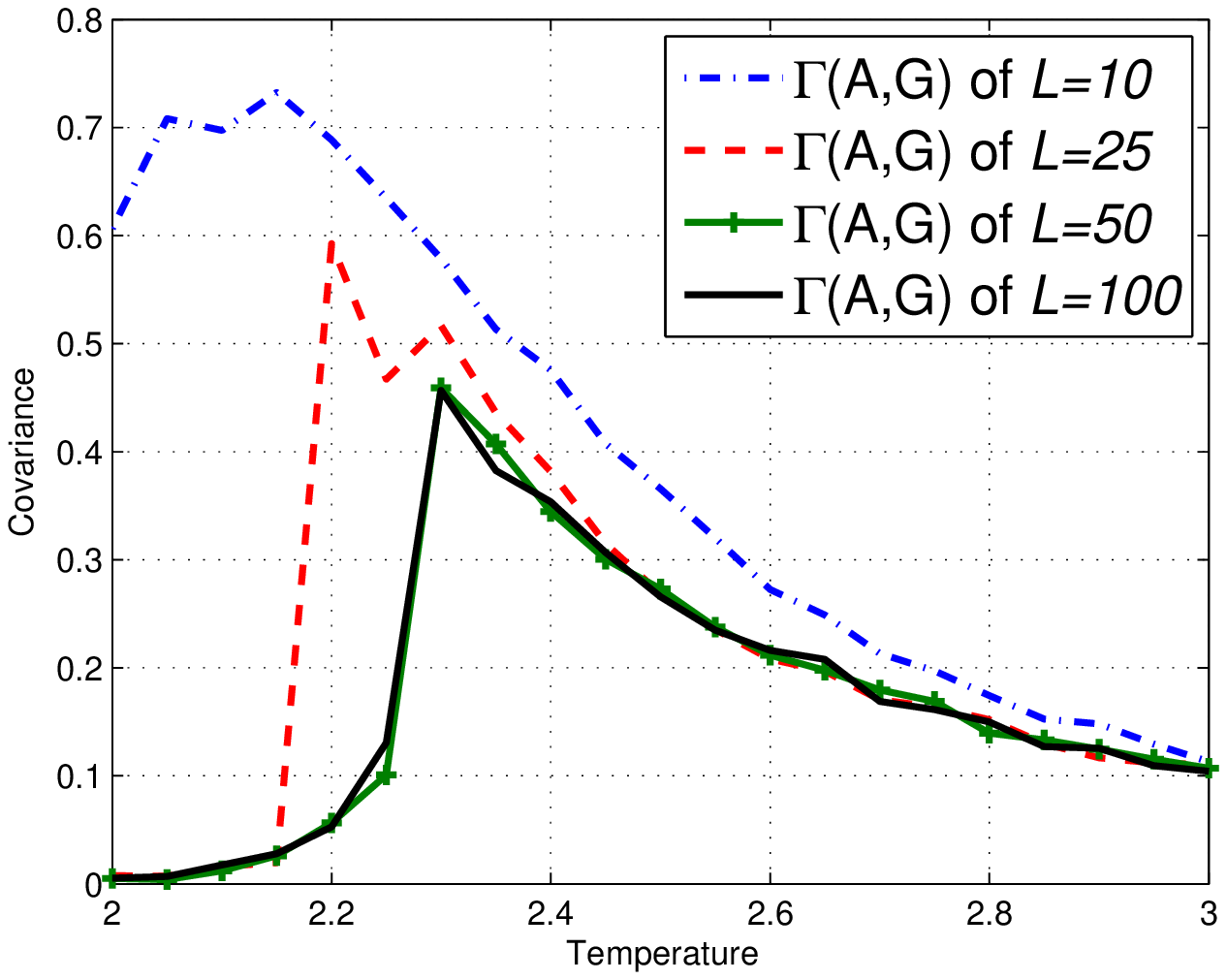}
\caption{\textbf{Covariance $\Gamma(A,G)$  on the \IM} with lengths $L=10,25,50,100$ obtained using equation (\ref{covising}).}
\label{cov100}
\end{minipage}
\end{figure}

In an infinite two dimensional lattice, the phase transition of the \IM\ with $J=1$ and $K_B=1$ is known to occur at the critical temperature $T_{c}=\frac{2}{log(1+\sqrt{2})} \approx 2.269185$ \cite{kim}. In a finite system, due to finite size effects, the critical values will not be quite as exact, we will call the temperature where the transition effectively occurs in the simulation as the crossover temperature $T_c$.  Susceptibility $\chi$ is an observable that is normally used to identify $T_c$ for the \IM\ as seen in Figure (\ref{suscep100}).  In order to define  $\chi$, let $m(n)=\sum_{i = 1}^{N} (s_i)_n$ be the sum of spins on a lattice of size $N$ at time steps $n=1, \cdots, S$.
The susceptibility~\cite{kim} is given by\begin{align}\label{susceptibility}
\chi=\frac{1}{TN} \left(E[m(n)^2]-E[m(n)]^2\right) 
\end{align}  
where $E[.]$ is the expectation in terms of temporal average and $T$ is temperature. The covariance on the \IM\ can be defined as
\begin{align}\label{covising}
\Gamma(X,Y)&=\Gamma(s_{X},s_{Y})=E_{}[s_Xs_{Y}]-E[s_X]E[s_Y]
\end{align}
where $X,Y \in \mathcal N$.
 
\begin{figure}[ht]
\begin{minipage}[b]{0.5\linewidth}
\centering
\includegraphics[width=\textwidth]{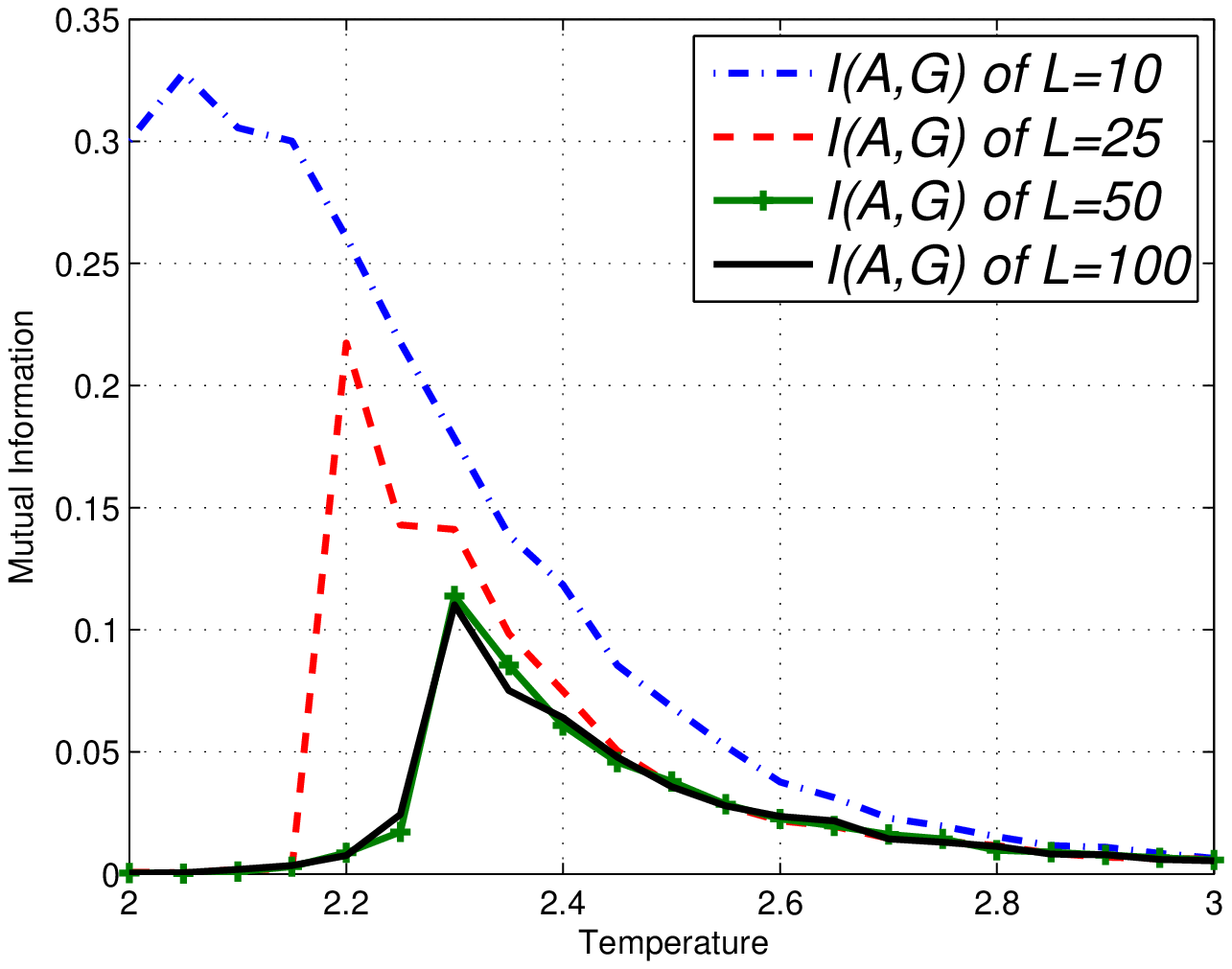}
\caption{\textbf{Mutual Information $I(A,G)$ on the \IM} with lengths $L=10,25,50,100$ obtained using equation (\ref{Isums}).}
\label{MI100}
\end{minipage}
\hspace{0.3cm}
\begin{minipage}[b]{0.5\linewidth}
\centering
\includegraphics[width=\textwidth]{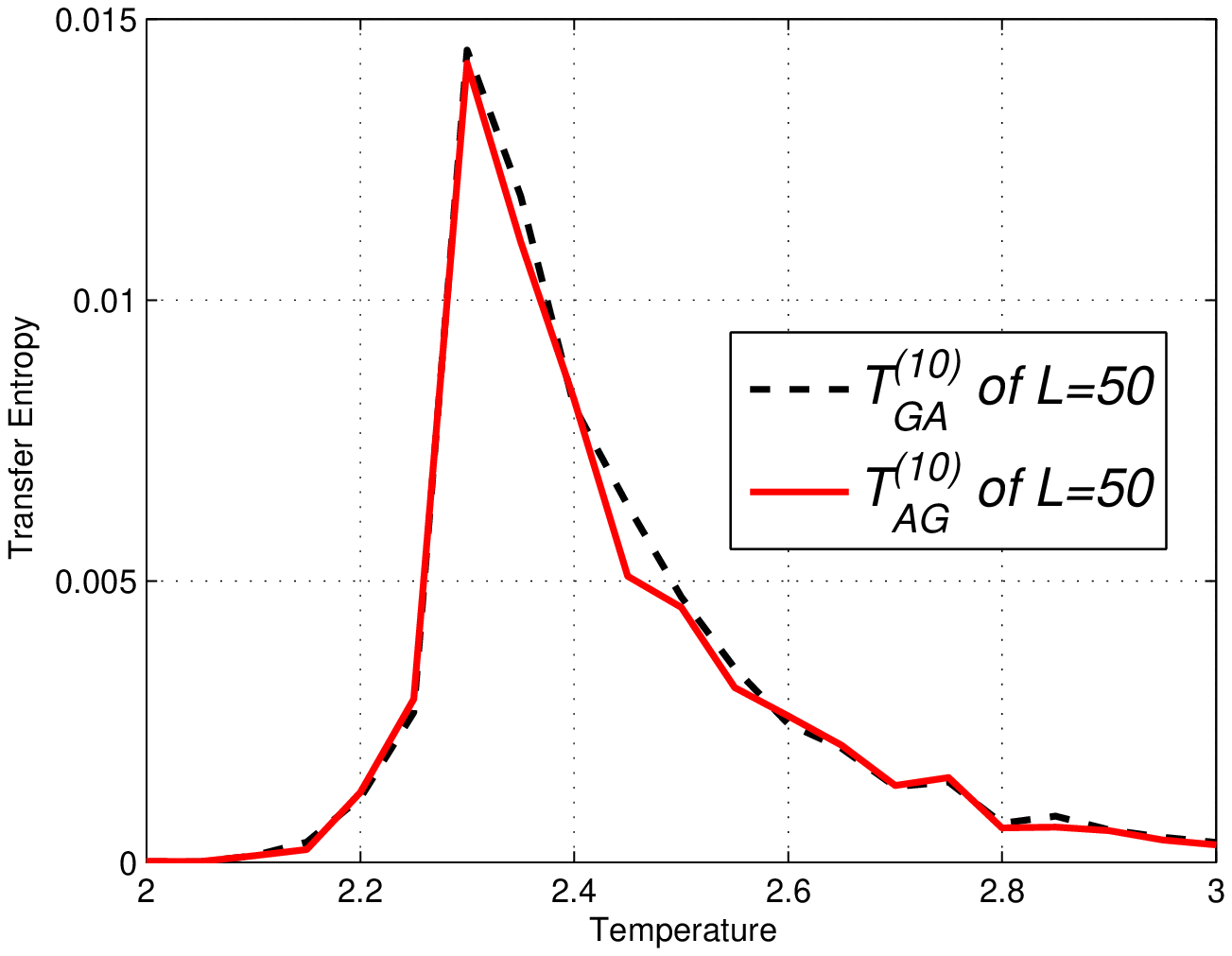}
\caption{\textbf{\TE\ $T^{(10)}_{AG}$ \&\ $T^{(10)}_{GA}$ on the \IM} of lengths $L=50$ obtained using equation (\ref{TEtau}). Peaks for both direction are at $T_c$.}
\label{TAGGA}
\end{minipage}
\end{figure}

To display measures applied on individual sites, let sites $A,B,G \in \mathcal N$ represent coordinates $[1,1]$ , $[2,2]$ and $[3,3]$ respectively. The values of the covariance $\Gamma(A,G)$ and $I(A,G)=I(s_{A},s_{G})$ is displayed in Figure (\ref{cov100}) and Figure (\ref{MI100}). It can be seen that for the \IM, \MI\ gives no more information than covariance. From this figure, one can see that the values are system size dependent up to system size $L=50$ or $N=2500$. We conclude from this that, up to this length scale correlations are detectable across the entire lattice\cite{kim}. Thus we shall frequently utilize $L=50$ when illustration is required.

Using time shifted variables we obtained the \TE\ $T_{YX}^{(\tau)}=T^{(\tau)}_{s_Y s_X}$ in Figure (\ref{ALLTEIM}).  One can see that there is no clear difference between $T^{(\tau)}_{GA}$ and $T^{(\tau)}_{AG}$ in the figures thus no direction of \cau\ can be established between $A$ and $G$. This is expected due to the symmetry of the lattice. More interestingly, the fact that   \TE\ peaks near $T_c$ can be due to the fact that at $T_c$ the  correlations span across the entire lattice. Therefore, one may say that the critical transition and collective behaviour in the \IM\ is detected by \TE\ as a type of \cau\ that is symmetric in both directions.
It is logical to interpret collective behaviour as a type of \cau\ in all directions since information is disseminated throughout the whole lattice when it is fully connected. This is an important fact to take into account when estimating \TE\ on \cs.

\section*{Amended \IM}

In the amended Ising model we introduce an explicit directed dependence between the sites $A$, $B$ and $G$ in order to study how well \TE\ is able to detect this causality. We will define the amended \IM\ using the algorithm outlined as follows. At each step in the algorithm a site chosen  at random will  be considered for flipping with a certain probability $\gamma_B$ except when $A$ or $B$ is selected where an extra condition needs to be fulfilled first before it can be allowed to change. If $(s_{G})_{n-t_{G}}=1$, $A$ (or $B$) can be considered for flipping with probability $\gamma_B$ as usual, however if $(s_{G})_{n-t_{G}}=-1$, no change is allowed. Thus only one state of $G$ ($s_{G}=1$ in this case) allows sites $A$ and $B$ to be considered for flipping. Therefore, although $A$ (and $B$) have their own dynamics, their changes still depend on $G$. 

\begin{figure}[ht]
\begin{minipage}[b]{0.5\linewidth}
\centering
\includegraphics[width=\textwidth]{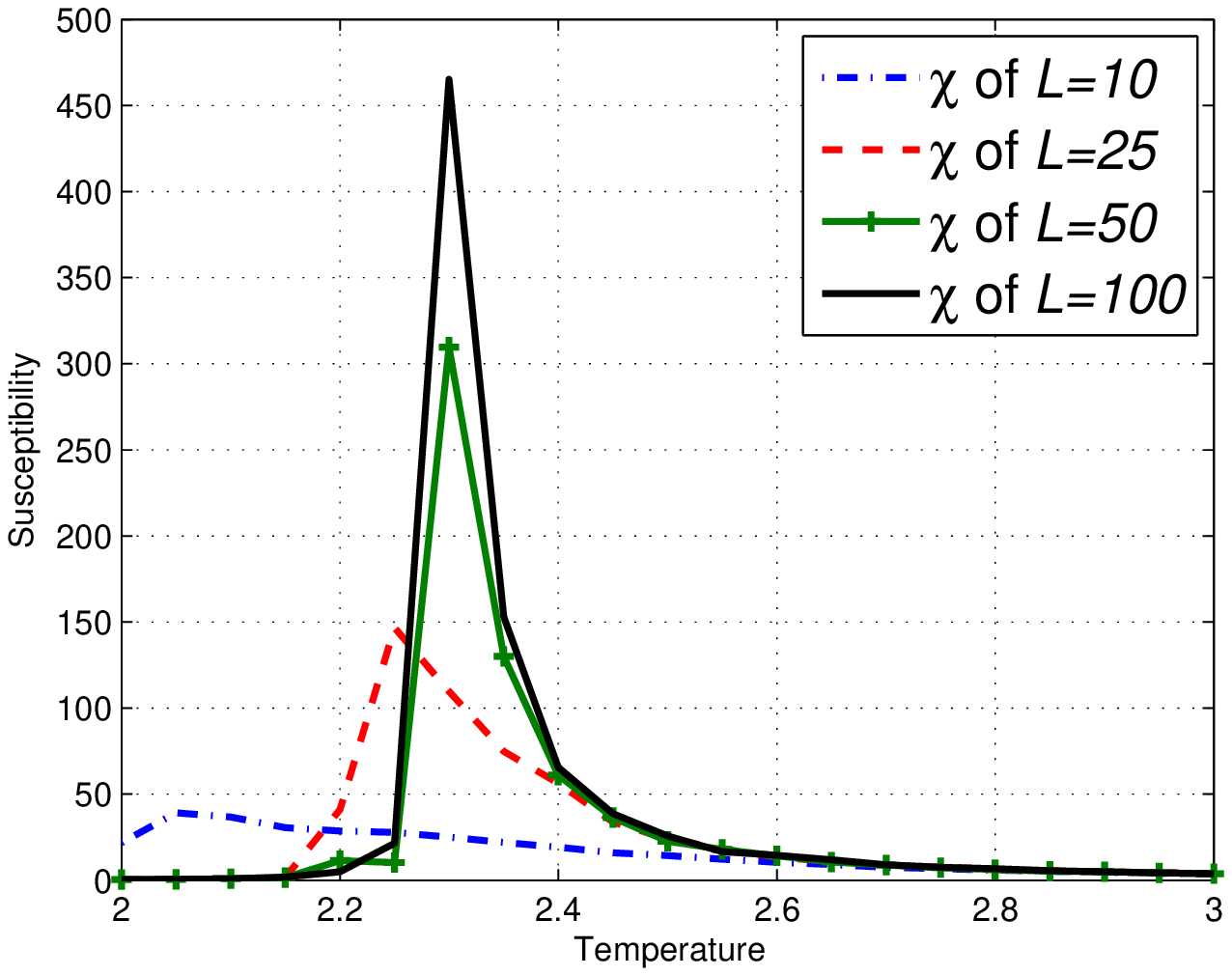}
\caption{\textbf{Susceptibility $\chi$ on the amended \IM} of lengths $L=10,25,50,100$ obtained using equation (\ref{susceptibility}). Peaks can be seen at respective $T_c$.}
\label{suscep100tau10}
\end{minipage}
\hspace{0.3cm}
\begin{minipage}[b]{0.5\linewidth}
\centering
\includegraphics[width=\textwidth]{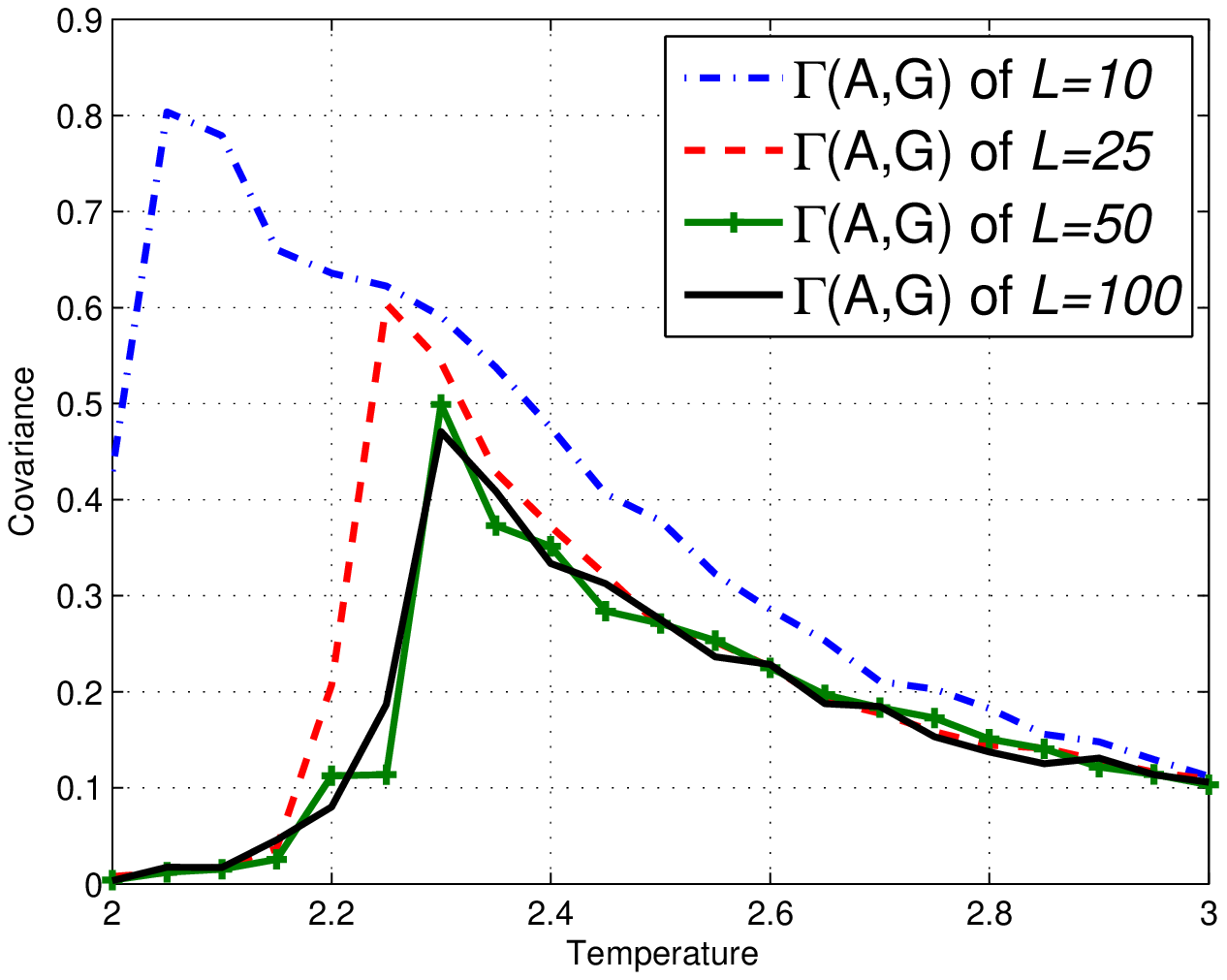}
\caption{\textbf{Covariance $\Gamma(A,G)$  on the amended \IM} of lengths $L=10,25,50,100$ obtained using equation (\ref{covising}). Peaks can be seen at respective $T_c$}
\label{cov100tau10}
\end{minipage}
\end{figure}

\begin{figure}[ht]
\begin{minipage}[b]{0.5\linewidth}
\centering
\includegraphics[width=\textwidth]{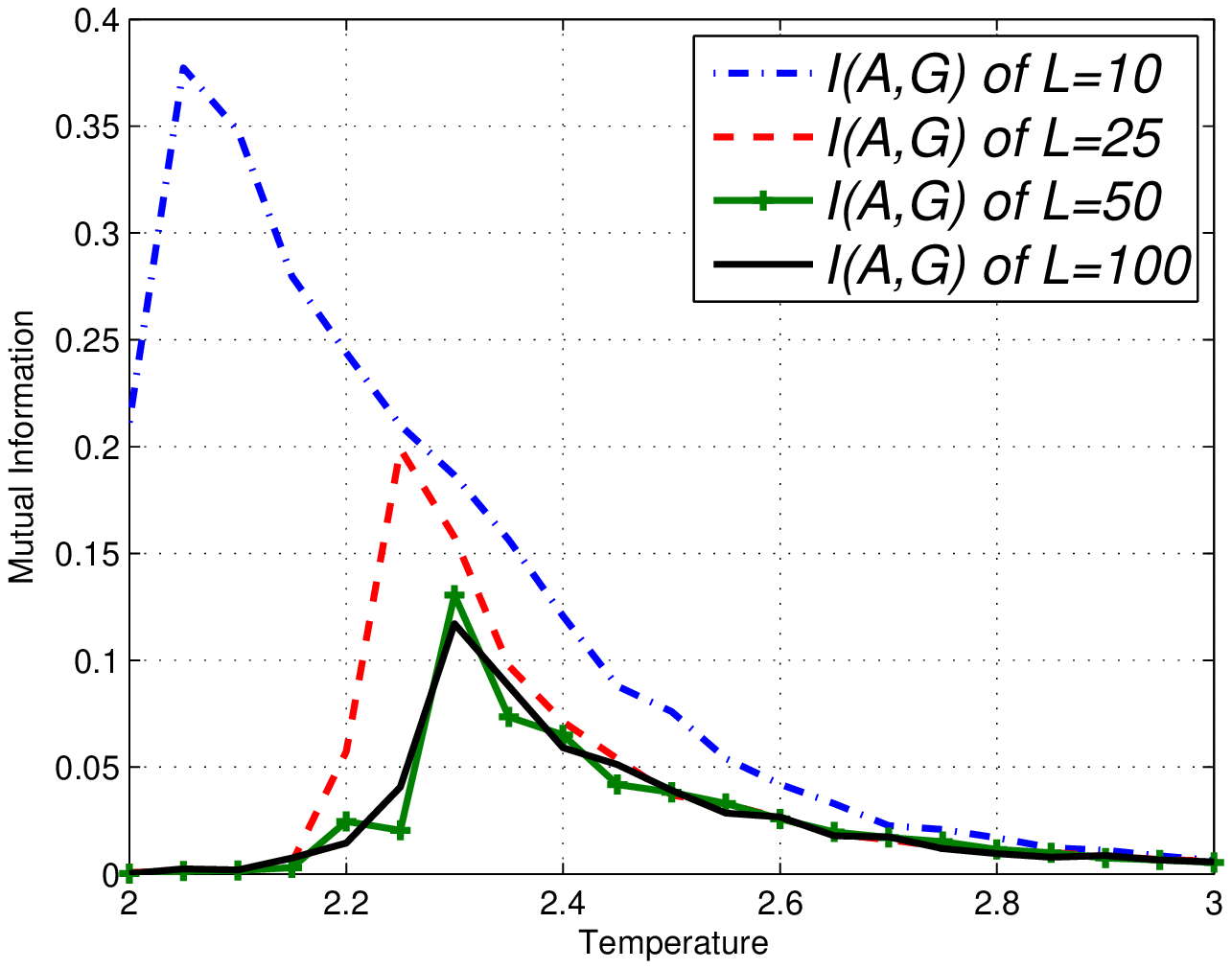}
\caption{\textbf{Mutual Information $I(A,G)$ on the amended \IM} with lengths $L=10,25,50,100$ obtained using equation (\ref{Isums}). Not much different from results on the \IM\ in Figure \ref{MI100}.}
\label{MI100tau10}
\end{minipage}
\hspace{0.3cm}
\begin{minipage}[b]{0.5\linewidth}
\centering
\includegraphics[width=\textwidth]{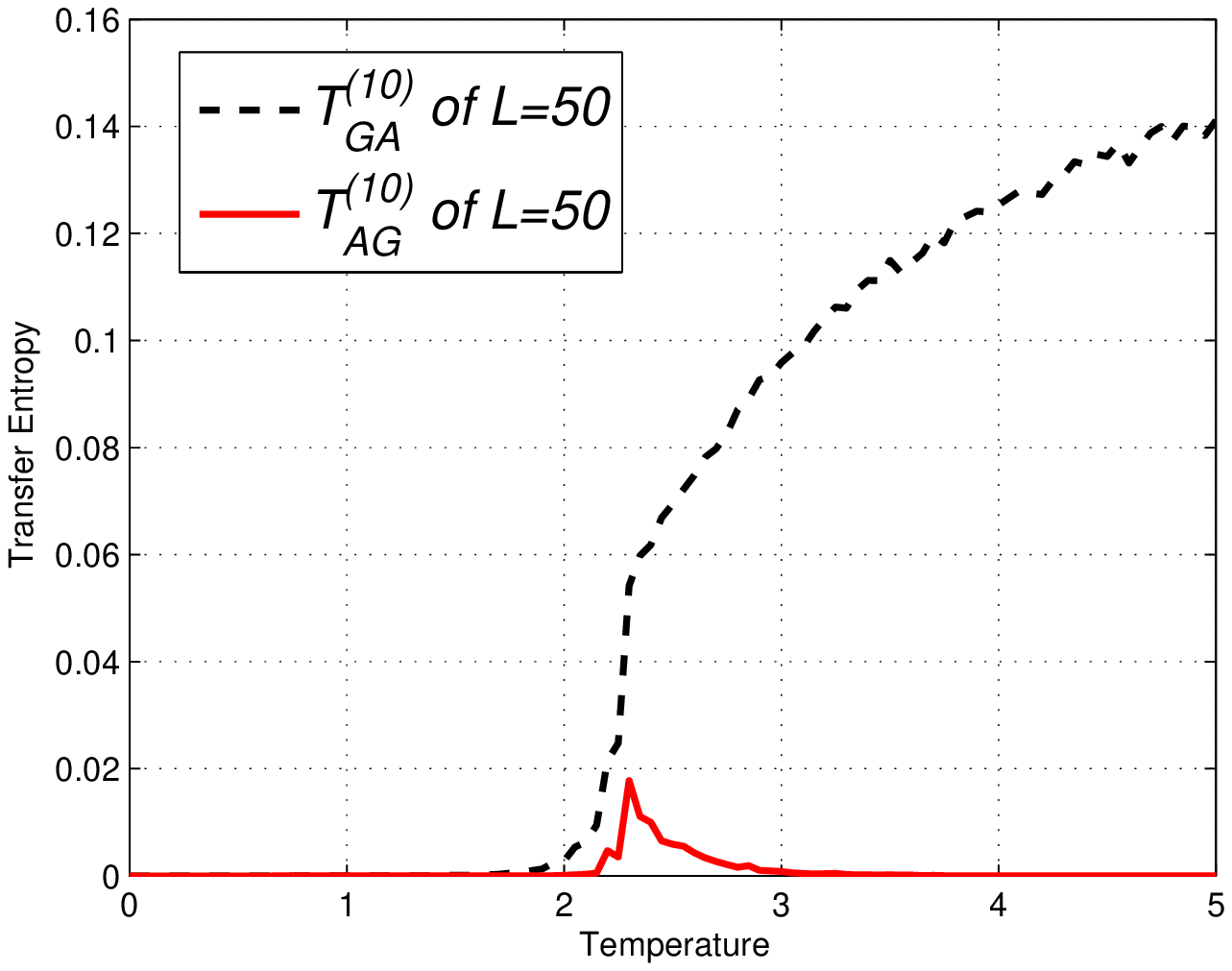}
\caption{\textbf{\TE\ $T^{(10)}_{AG}$ \&\ $T^{(10)}_{GA}$ on the amended \IM} of lengths $L=50$ and $t_G=10$, obtained using equation (\ref{TEtau}). Direction $G \rightarrow A$ at time lag $10$ is indicated. Very different from result on \IM\ in Figure \ref{TAGGA}.}
\label{TAGGA25}
\end{minipage}
\end{figure}

\begin{figure}[ht]
\begin{minipage}[b]{0.5\linewidth}
\centering
\includegraphics[width=\textwidth]{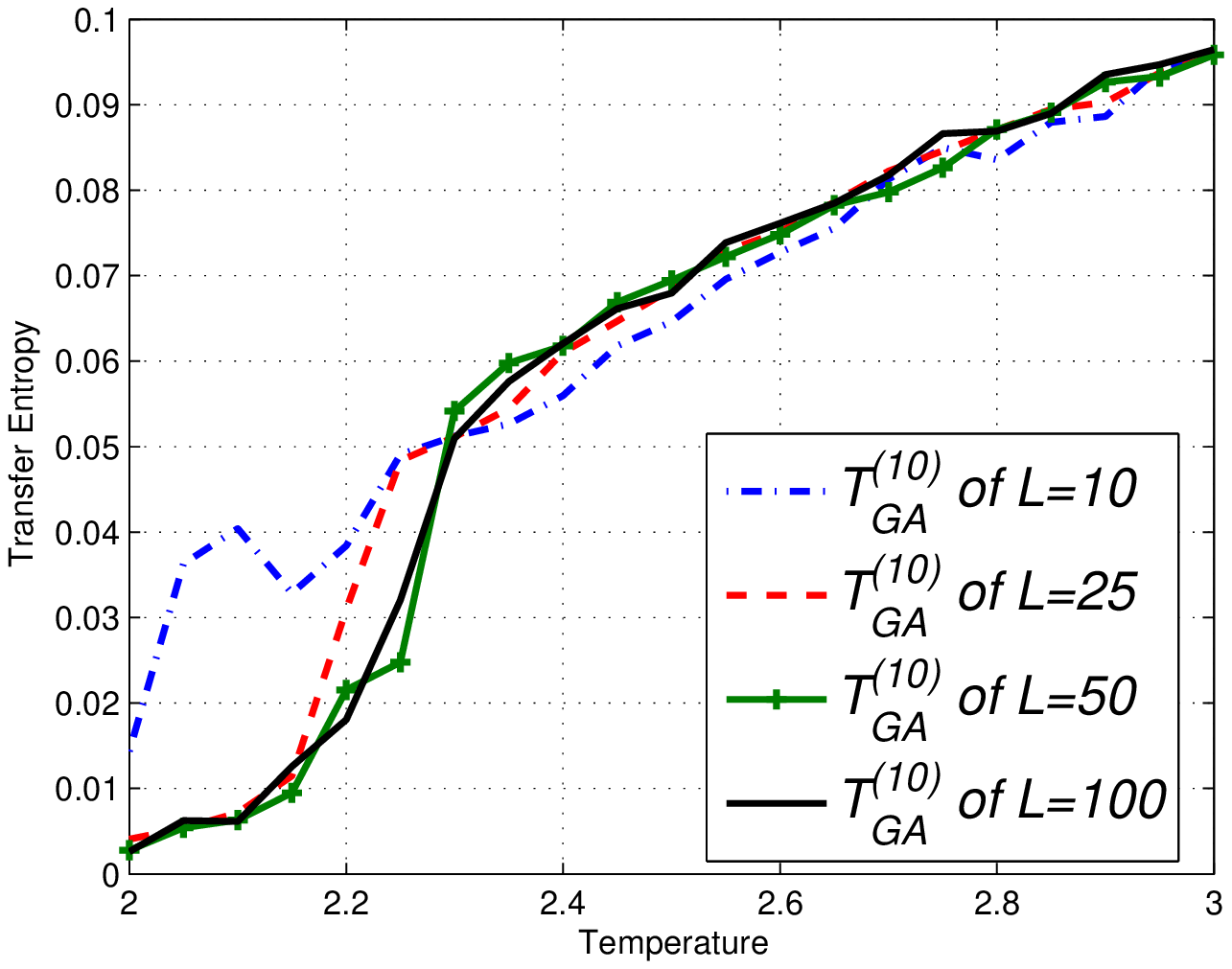}
\caption{\textbf{\TE\ $T^{(10)}_{GA}$ on the \IM} of lengths $L=10,25,50,100$ obtained using equation (\ref{TEtau}). Values continue to increase after $T_c$.}
\label{TEGA100tau10}
\end{minipage}
\hspace{0.3cm}
\begin{minipage}[b]{0.5\linewidth}
\centering
\includegraphics[width=\textwidth]{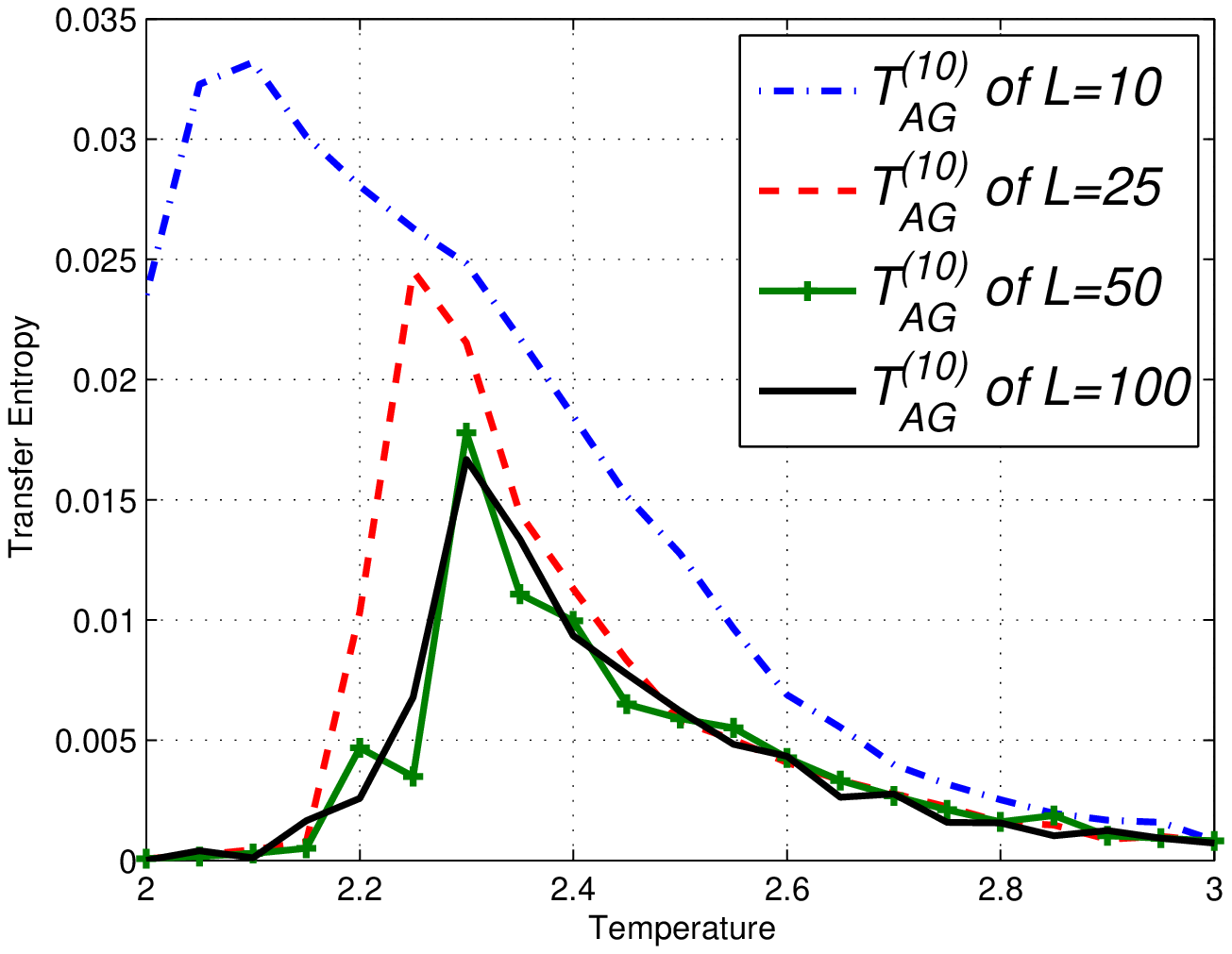}
\caption{\textbf{\TE\ $T^{(10)}_{AG}$ on the \IM} of lengths $L=10,25,50,100$ obtained using equation (\ref{TEtau}). Peaks can be seen at respective $T_c$ similar to \IM\ results.}
\label{TEAG100tau10}
\end{minipage}
\end{figure}

We simulated the amended \IM\ with $t_G=10$ for different lattice lengths  $L$.  Figures (\ref{suscep100tau10}) display the values of susceptibility $\chi$ on the model and the peaks clearly show the presence of $T_c$  in our model.   Figures (\ref{cov100tau10}) and (\ref{MI100tau10}) display the values of  the covariance $\Gamma(A,G)$ and the \MI\ $I(A,G)$.  We reiterate that our correlations reach across the system for $L\le50$ \cite{witthauer,kim}. While covariance and \MI\ gives similar results to those of the standard \IM, a difference is clearly seen in \TE\ values. Figure (\ref{TAGGA25}-\ref{TEAG100tau10}) displays the contrasts of $T^{(10)}_{AG}$ and $T^{(10)}_{GA}$ on the amended \IM\ which explicitly indicates the direction of \cau\ $G \rightarrow A$.

\begin{figure}[ht]
\begin{minipage}[b]{0.5\linewidth}
\centering
\includegraphics[width=\textwidth]{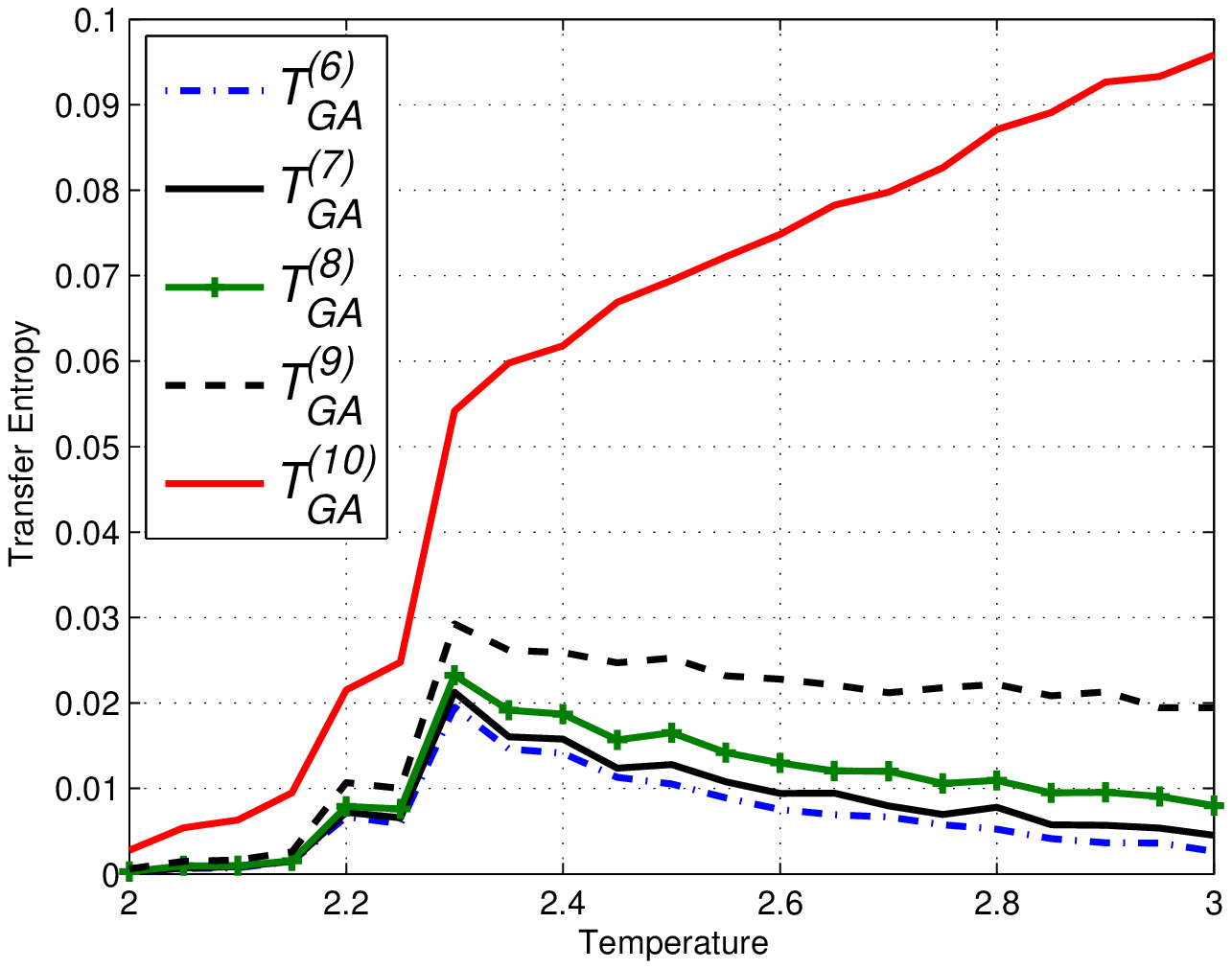}
\caption{$T^{(\tau)}_{GA}$ versus $T$ for different time lags $\tau$ in amended \IM\ with $t_G=10$ and $L=50$ using equation (\ref{TEtau}). }
\label{TEtau10L50}
\end{minipage}
\hspace{0.2cm}
\begin{minipage}[b]{0.5\linewidth}
\centering
\includegraphics[width=\textwidth]{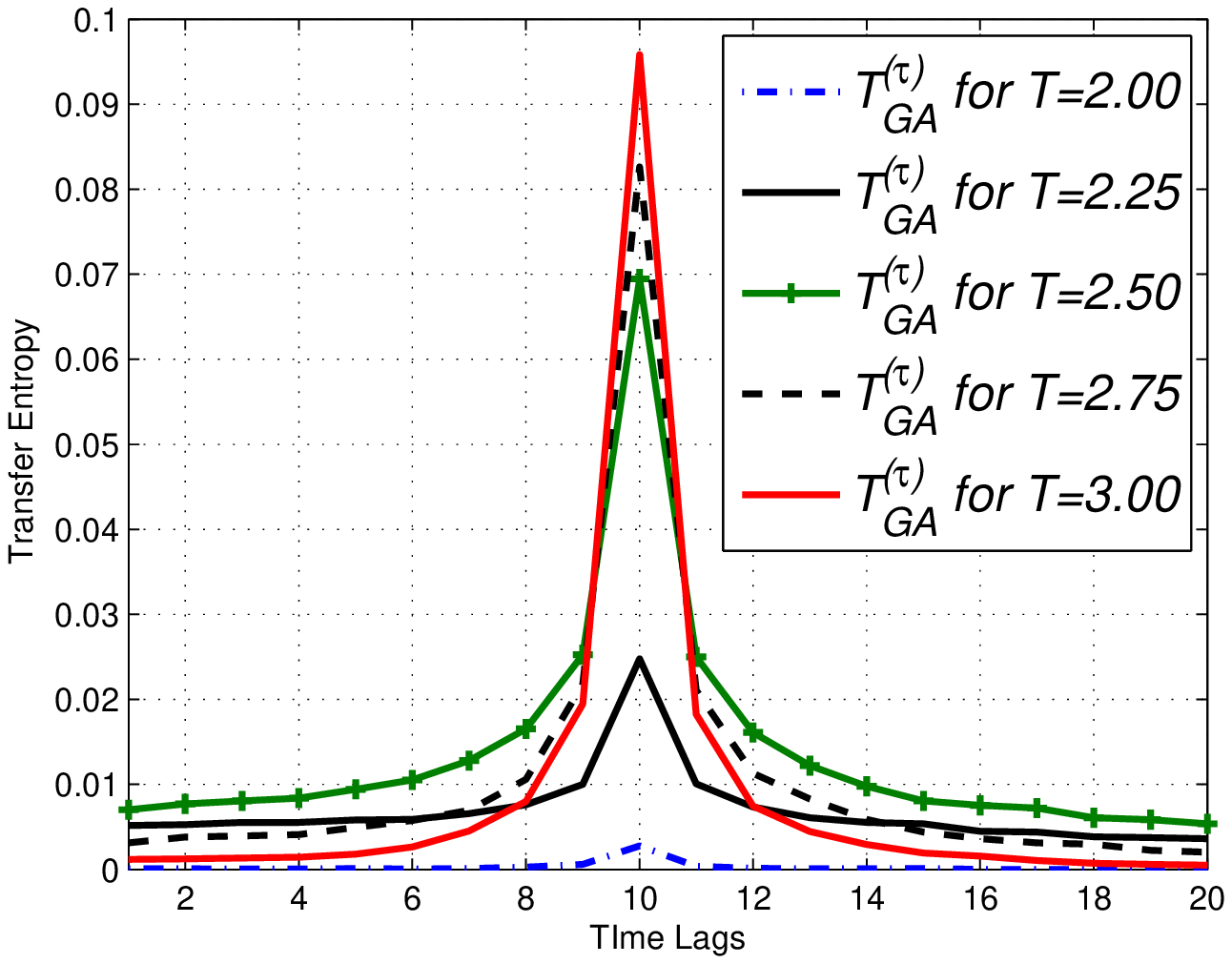}
\caption{$T^{(\tau)}_{GA}$ versus  $\tau$ for different temperatures $T$ in amended \IM\ with $t_G=10$ and $L=50$ using equation (\ref{TEtau}). $T_c \approx 2.3$.}
\label{TEsvtauAIM}
\end{minipage}
\end{figure}

The effect of deviation from the predetermined causal lag $t_G=10$, can be clearly seen in Figure (\ref{TEtau10L50}), where for the values of $T^{(\tau)}_{GA}, \tau \ne 10$ reduces to $0$ but at different rates depending on the deviation of $\tau$ from $t_G$. The further away from $t_G$, the faster the decrease to $0$. Figure (\ref{TEsvtauAIM}) is simply Figure (\ref{TEtau10L50}) plotted over different time lags $\tau$ to  illustrate how \TE\  correctly and distinctly identified causal lag $t_G=10$.
  
\begin{figure}[ht]
\begin{minipage}[b]{0.33\linewidth}
\centering
\includegraphics[width=\textwidth]{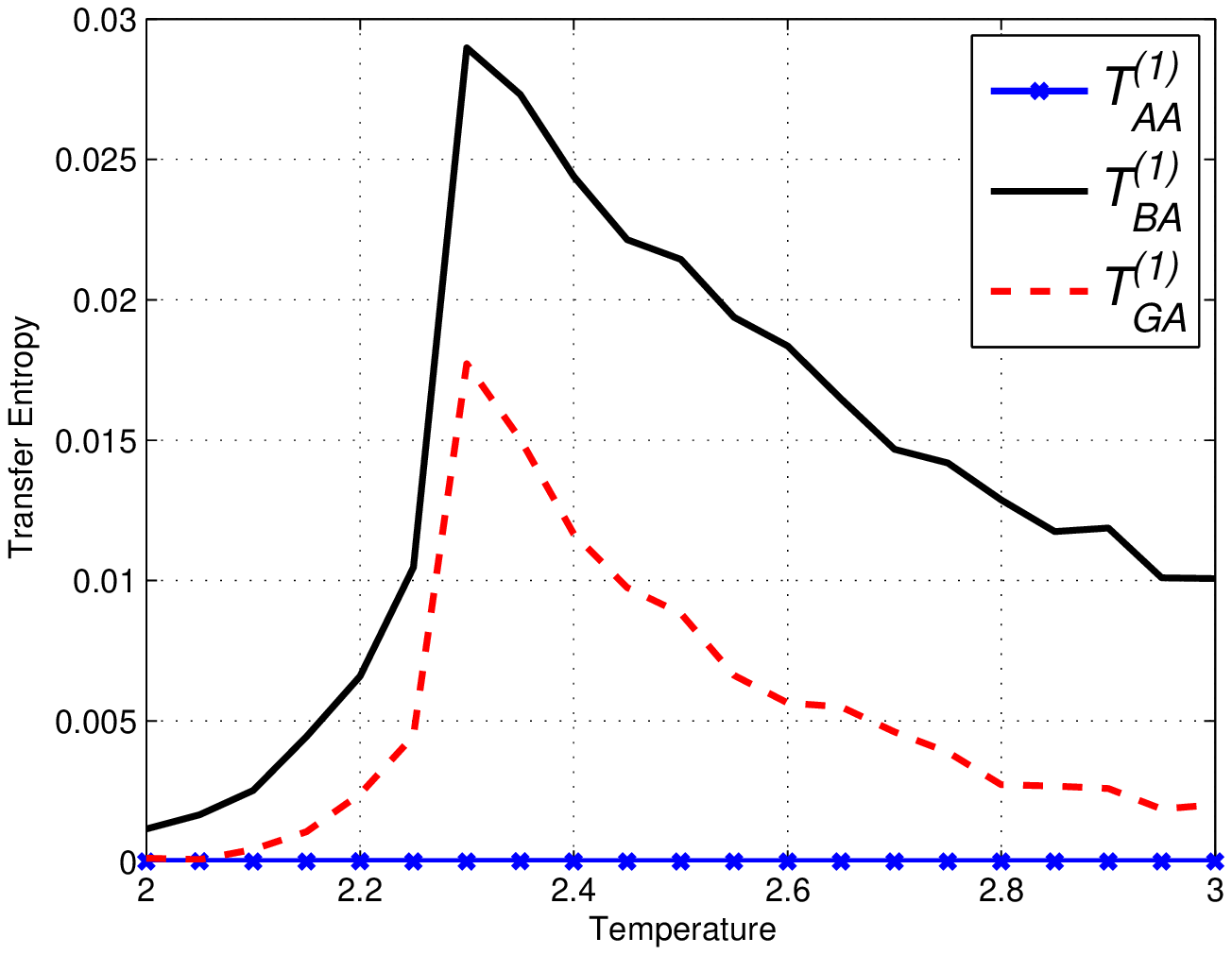}
\caption{$T^{(1)}_{AA}$, $T^{(1)}_{BA}$ and $T^{(1)}_{GA}$ in the \IM\ with $L=50$. $T^{(1)}_{BA} > T^{(1)}_{GA}$ due to distance in space.}
\label{TEwithAbasicL50} 
\end{minipage}
\hspace{0.05cm}
\begin{minipage}[b]{0.33\linewidth}
\centering
\includegraphics[width=\textwidth]{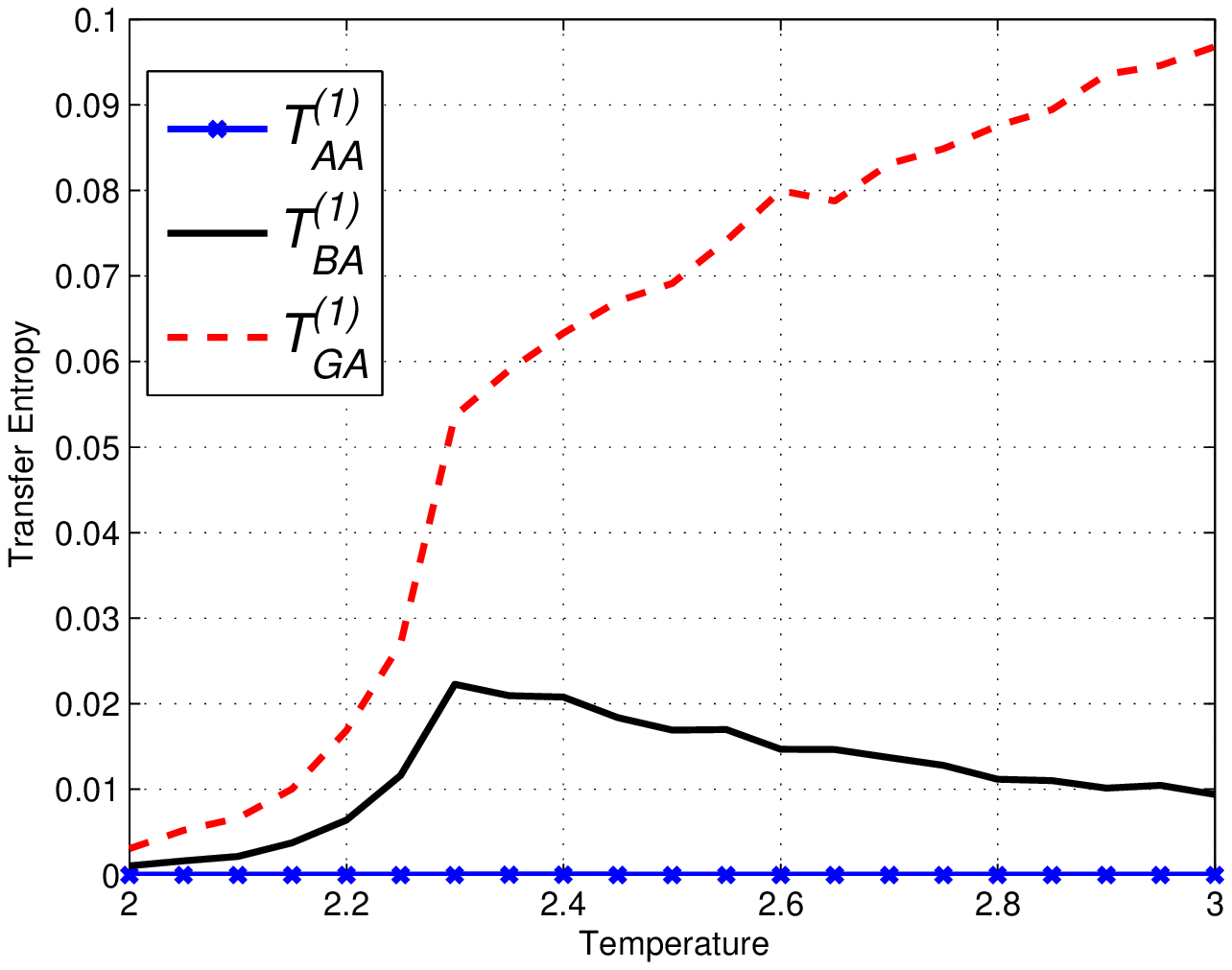}
\caption{$T^{(1)}_{AA}$, $T^{(1)}_{BA}$ and $T^{(1)}_{GA}$ in the amended \IM\ with $L=50$ and $t_G=1$. $T^{(1)}_{BA} < T^{(1)}_{GA}$ due to implanted `causal' lag.}
\label{TEwithAL50} 
\end{minipage}
\hspace{0.05cm}
\begin{minipage}[b]{0.33\linewidth} 
\centering
\includegraphics[width=\textwidth]{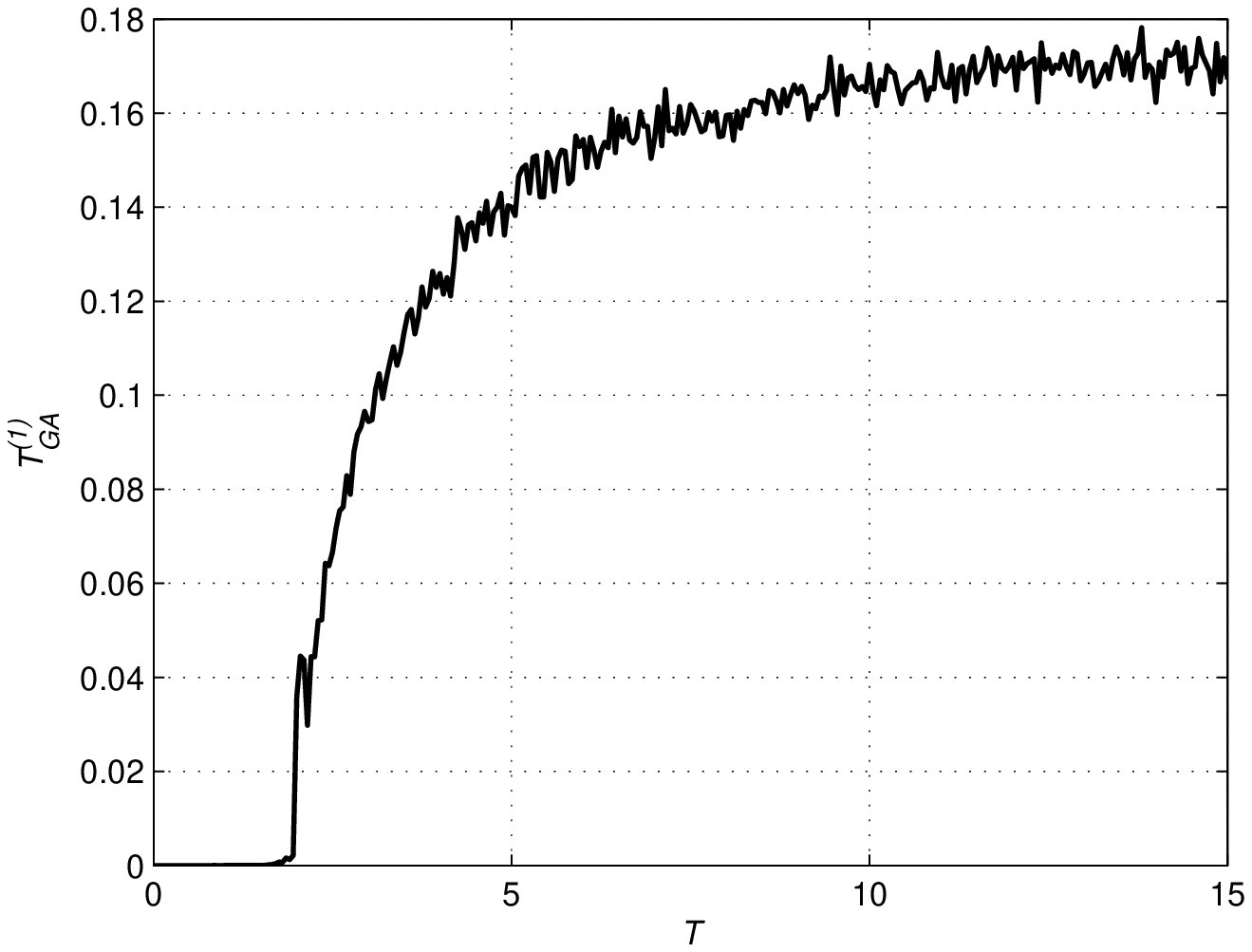}
\caption{$T^{(1)}_{GA}$ in Figure \ref{TEwithAL50} up to $T=15$. \TE\ in the indicated direction stabilizes at higher temperature.}
\label{temp0to15} 
\end{minipage}
\end{figure}

That temperature is a main factor in influencing the strength of \TE\ values is apparent in all the figures in this section. One can observe that the \TE\ values approaches $0$ as they get further away from $T_c$ except when the time lag matches the delay  induced by definition between the dynamics of the two spins $A$ and $B$   and the $G$ spin, in which case the \TE\ value stabilizes to a certain fixed value as seen in Figure (\ref{temp0to15}).  In the vicinity of $T_c$, the lattice is highly correlated thus subsequently leading to higher  values of \TE. The increase and value stabilization after $T_c$ is due to the fact that, as temperature increases, the probability for all spin flips approaches a uniform distribution. This leads to transfer of information between site $G$ and sites $A$ and $B$ occurring much more frequently at elevated temperature.

Figure (\ref{TEwithAbasicL50}) and (\ref{TEwithAL50}) display \TE\ values for the \IM\ and amended \IM\ with $t_G=1$ respectively. The figures illustrate the mechanism in which \TE\ detects the predefined causal delay. Consider  the following question: which site `causes' site $A$? Firstly we see that $T^{(1)}_{AA}$ is zero in both figures due to the definition in equation (\ref{TEtau}). Note that this is only for $\tau=1$, if $\tau \ne 1$ the \TE\ value will be nonzero and also peak at $T_c$. More importantly  we see that $T^{(1)}_{GA}$ is different from $T^{(1)}_{BA}$. In Figure (\ref{TEwithAbasicL50}) the difference is due to distance in space and nearest neighbour interaction in the model, thus $T_{GA}^{(1)} < T_{BA}^{(1)}$ since $G$ is further away from $A$ than $B$. 
 But in Figure (\ref{TEwithAL50}), the opposite is true and distance in space does not dominate in this interaction. The figure very clearly indicates that $G$ `causes' $A$ at $\tau=1$ and $B$ does not. In other words, in the amended \IM\ \TE\ identifies $G$ as a source in which one of the target is $A$, whereas in the \IM\ the expected nearest neighbour dynamics presides. This result is only obtained for measures sensitive to transition probabilities. Measures that depend only on static probabilities such as covariance, \MI\ and  \CMI\ will only give values in accordance to the underlying nearest neighbour dynamics in both the \IM\ and the amended \IM \cite{fatimah}.

\subsection*{\TE, directionality and change}

\begin{figure}[ht]
\begin{minipage}[b]{0.33\linewidth}
\centering
\includegraphics[width=\textwidth]{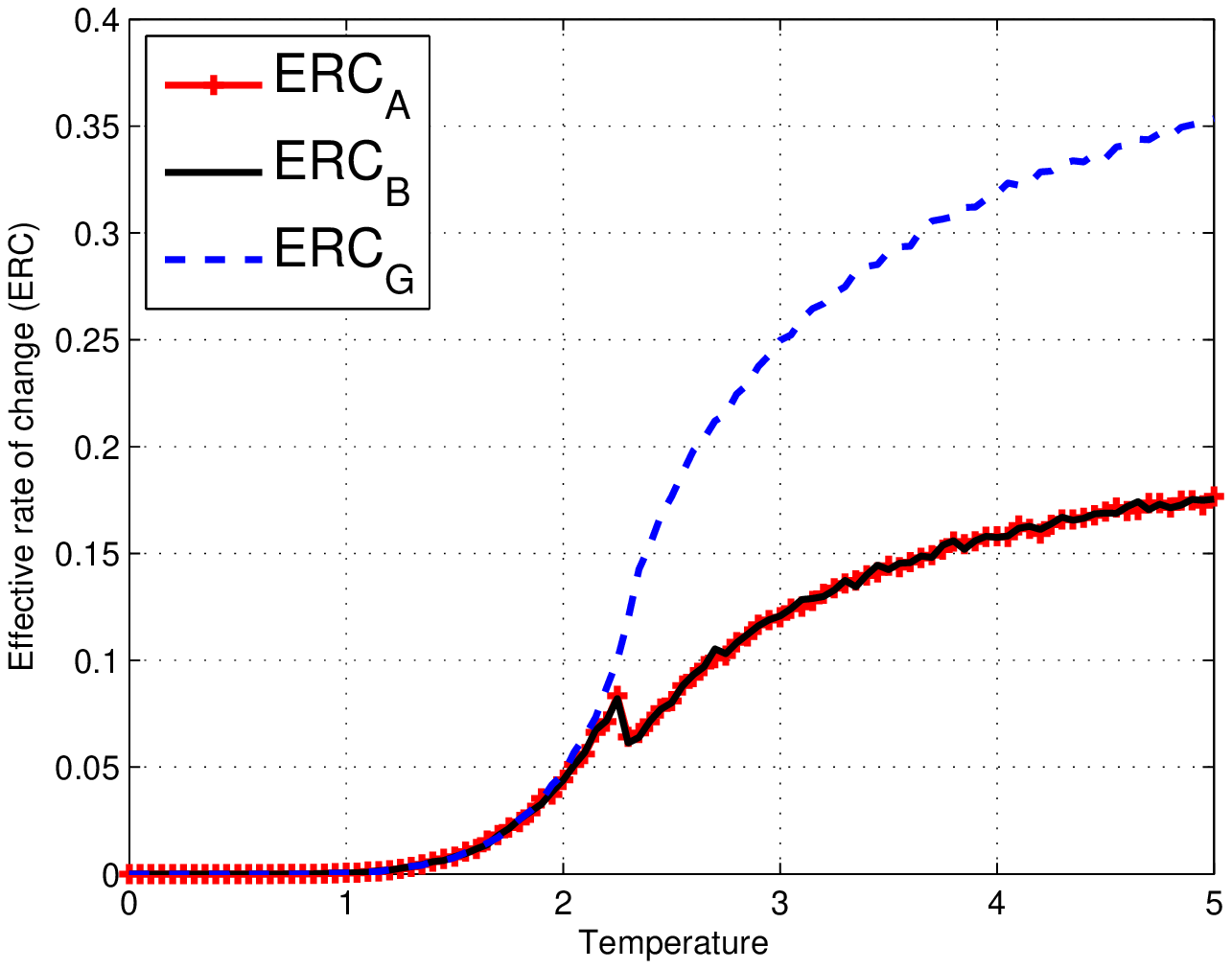}
\caption{$ERC$ of sites $A$, $B$ and $G$ on amended \IM\ with $t_G=10$ and $L=50$.}
\label{ercL50} 
\end{minipage}
\hspace{0.05cm}
\begin{minipage}[b]{0.33\linewidth}
\centering
\includegraphics[width=\textwidth]{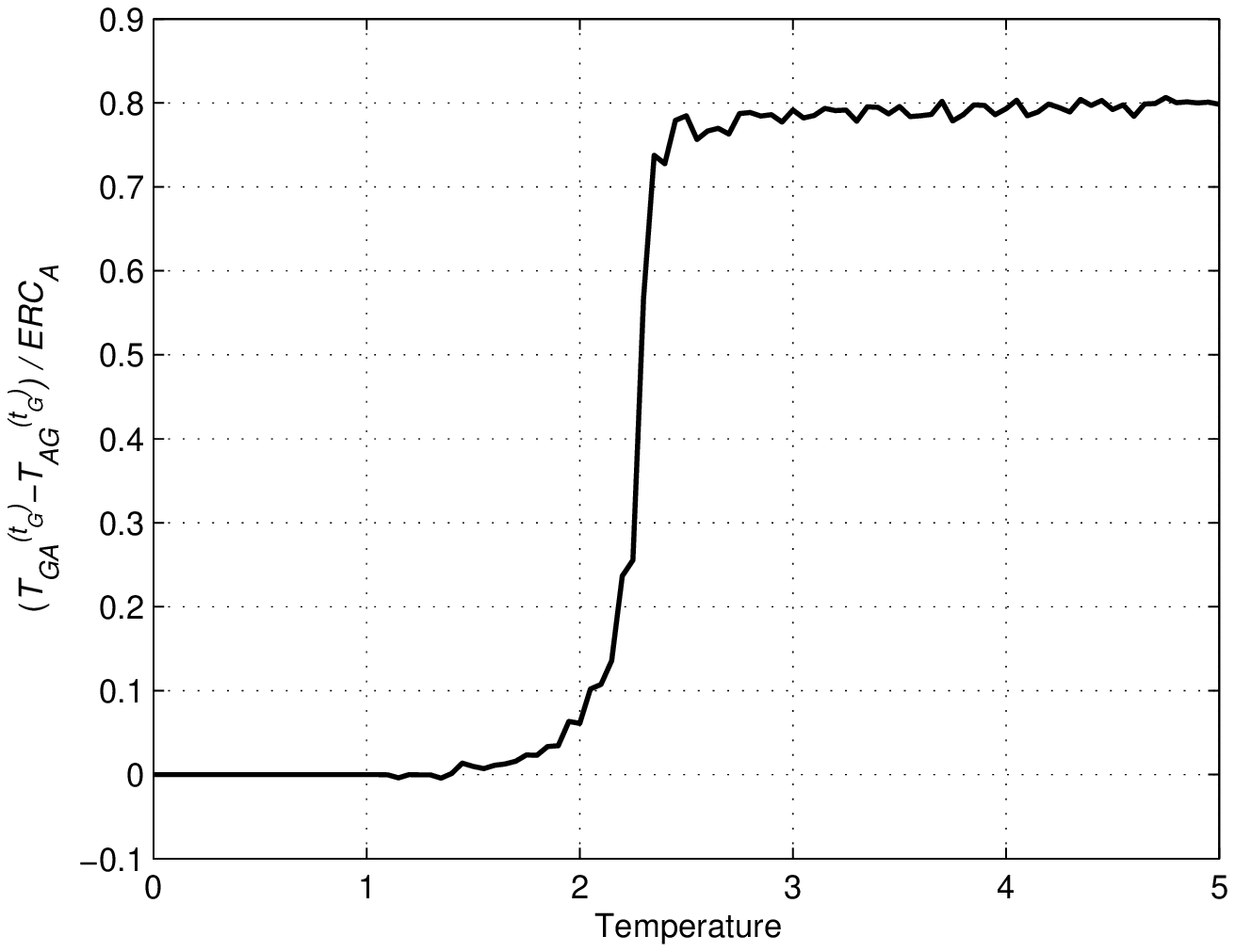}
\caption{$\frac{T^{(t_G)}_{GA}-T^{(t_G)}_{AG}}{ERC_A}$ on amended \IM\ with $t_G=10$ and $L=50$}
\label{dtL25} 
\end{minipage}
\hspace{0.05cm}
\begin{minipage}[b]{0.33\linewidth} 
\centering
\includegraphics[width=\textwidth]{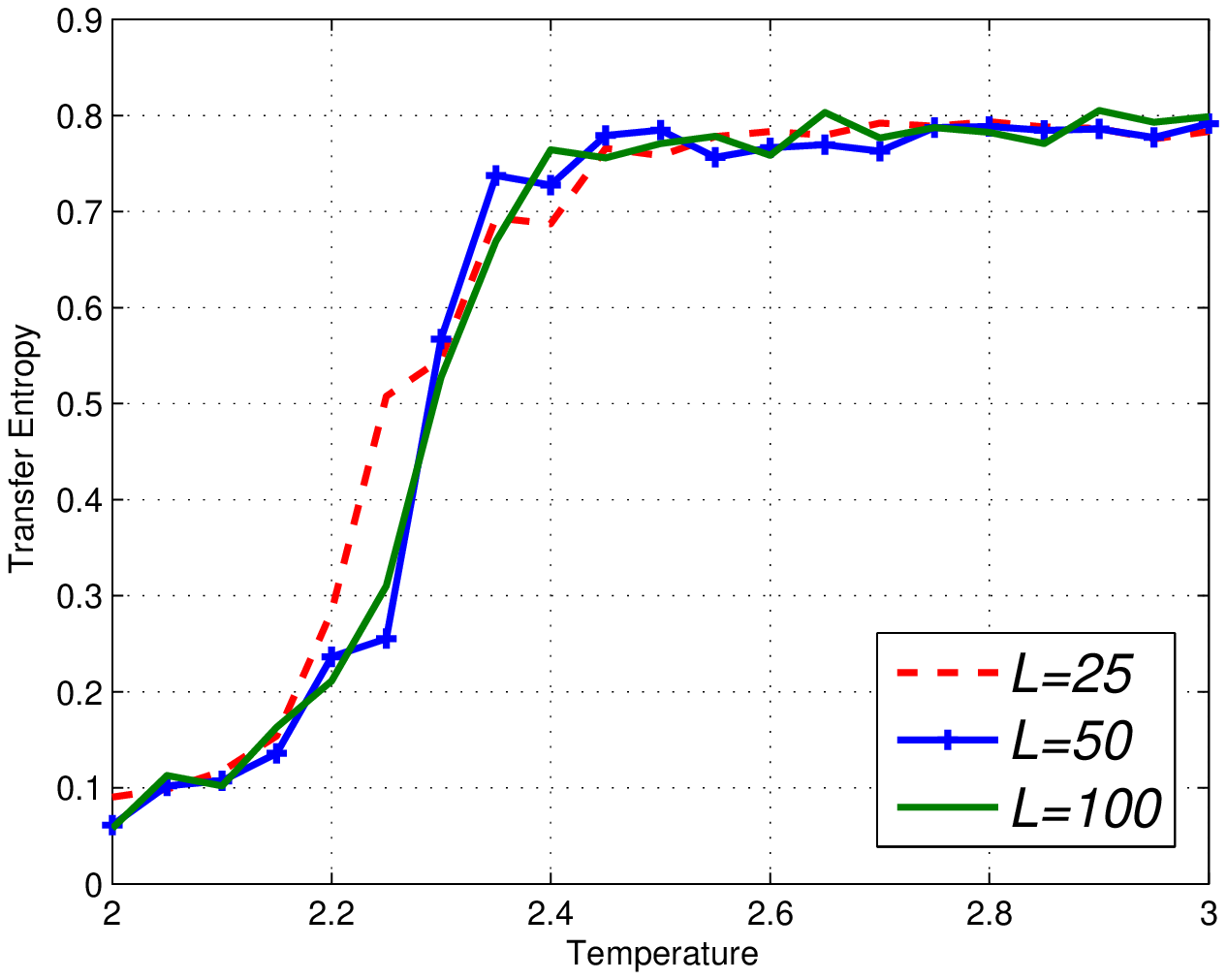}
\caption{
$\frac{T^{(t_G)}_{GA}-T^{(t_G)}_{AG}}{ERC_A}$ on amended \IM\ with $t_G=10$ and $L=25, 50, 100$.}
\label{DT} 
\end{minipage}
\end{figure}

In order to understand the dynamics of of each site  we calculate the effective rate of change (ERC) in relation to the transition probabilities. Let $ERC_X=P(X_n \ne X_{n-1})$ for any site $X$ on the lattice. Figure (\ref{ercL50}) illustrates how $ERC_A$ and $ERC_B$ are equal, as expected, and significantly different from $ERC_G$. In Figure (\ref{TAGGA25}), the corresponding \TE\  in both directions are displayed. At higher temperatures, it can be clearly seen that $T^{(t_G)}_{GA}$ is larger than $T^{(t_G)}_{AG}$. However for temperatures near $T_c$ it is not as clear and therefore to highlight the relative values we calculate $\frac{T^{(t_G)}_{GA}-T^{(t_G)}_{AG}}{ERC_A}$ in Figure (\ref{dtL25}) and Figure (\ref{DT}) where $\frac{T^{(t_G)}_{GA}-T^{(t_G)}_{AG}}{ERC_A} =0$ if  $ERC_{A} = 0$. We see that this value actually gives a clear jump at $T_c$ and remains more or less a constant after $T_c$. Therefore even though \TE\ in neither direction is zero, a clear indication of directionality can be obtained. Interestingly, the division with ERC brought out the clear phase transition-like behaviour that seems to distinguish the situation below and above $T_c$. Referring back to Figure (\ref{TAGGA}) of the unamended \IM\, we can clearly see that $\frac{T^{(t_G)}_{GA}-T^{(t_G)}_{AG}}{ERC_A} \approx 0$ for any direction in the unamended \IM. We have demonstrated that $\frac{T^{(t_G)}_{GA}-T^{(t_G)}_{AG}}{ERC_A}$ is able to  cancel out the symmetric contribution from the  collective behaviour and only captures the imposed directed interdependence. 

In his introductory paper \cite{sch00}, Schreiber warns that in certain situations due to different information content as well as different information rates, the difference in magnitude should not be relied on to imply directionality unless \TE\ in one direction is $0$. We have shown that when collective behaviour is present on the \IM, the value of \TE\ cannot possibly be $0$. We suggest that this is due to fact that collective behaviour is as a type of \cau\ (disseminating information in all directions) and thus the \TE\ is correctly indicating `cause' in all directions. The clear difference in \TE\ magnitude (even at $T_c$) observed when the model is amended indicates that the {\it difference} in \TE\ can indeed serve as an indicator of directionality in systems with emergent cooperative behaviour. We have seen that \TE\ is influenced by the nearest neighbour interactions, collective behaviour and the ERC. In the next section we use the Random Transition model to further investigate how the ERC influences the \TE. 

\section*{Random Transition Model}

In the amended \IM\ we implemented a causal lag as a restriction of one variable on another, in a way that a value of the source variable will affect the possible changes of the target variable. It is this novel concept of implementing \cau\ that we will analyze and expand in the Random Transition model. Let $\mu_X$, $\mu_Y$ and $\mu_Z$, be the independent probabilities for the stochastic swaps of  the variables $X$, $Y$ and $Z$ at every time step respectively. In addition to that, a restriction is placed on $X$ and $Y$  such that they are only allowed to do the stochastic swap with probability $\mu_X$ and $\mu_Y$ if the state of  $Z_{n-t_Z}$ fulfills a certain condition. This restriction means  that $X$ and $Y$ can only change states if $Z$  is in the conditioned state at time step $n-t_Z$ thus creating a `dependence' on $Z$, analogous to the dependence of $A$ and $B$ on $G$ in the amended \IM. However in this model we allow the number of states $n_s$ to be more than just two. The purpose of this is twofold, on one hand it contributes towards verifying that the behaviours of \TE\ observed on the amended \IM\ does extend to cases where $n_s>2$. On the other hand, the model also serves to highlight different properties of \TE\ as well as the very crucial issue of probability estimation that may lead to misleading results.  The processes are initialized randomly and independently. The swapping probabilities are taken to be $\mu_X=\mu_Y=\mu_Z=\frac{1}{n_s}$, thus enabling \TE\ values to be calculated analytically (see appendix for detailed analytic formulations).

The unclear meaning of the magnitude of \TE\ is one of its main criticism~\cite{pompe11,runge12b}. This is partly due to the ERC which incorporates both external and internal influences, the separation of which is rather unclear. The advantage of investigating \TE\ on the Random Transition model is that the ERC can be defined in terms of internal and external elements i.e. for any variable $X$ we have that
$$ERC_{X}=P(X_n\neq X_{n-1})=\sum_{\beta \ne \alpha} P(X_n=\alpha|X_{n-1}=\beta)=\mu_X\Omega,$$
where $\mu_X$ is the internal transition probability of $X$ and $\Omega$ represents the external influence applied on $X$. If the condition in our model is that $Z_{n-1}=1$ for $X_n$ and $Y_n$ to change values then,  $\Omega=P( \text{ condition fulfilled })=P(Z_{n-1}=1)$ so that 
$
ERC_X=\mu_XP(Z_{n-1}=1)
$
and
$
ERC_Y=\mu_YP(Z_{n-1}=1)$. However, for the source $Z$ which has no external influence, $\Omega=1$ and consequently $
ERC_Z=P(Z_n \ne Z_{n-1})=\mu_Z.
$

\begin{figure}[ht]
\begin{minipage}[b]{0.5\linewidth}
\centering
\includegraphics[width=\textwidth]{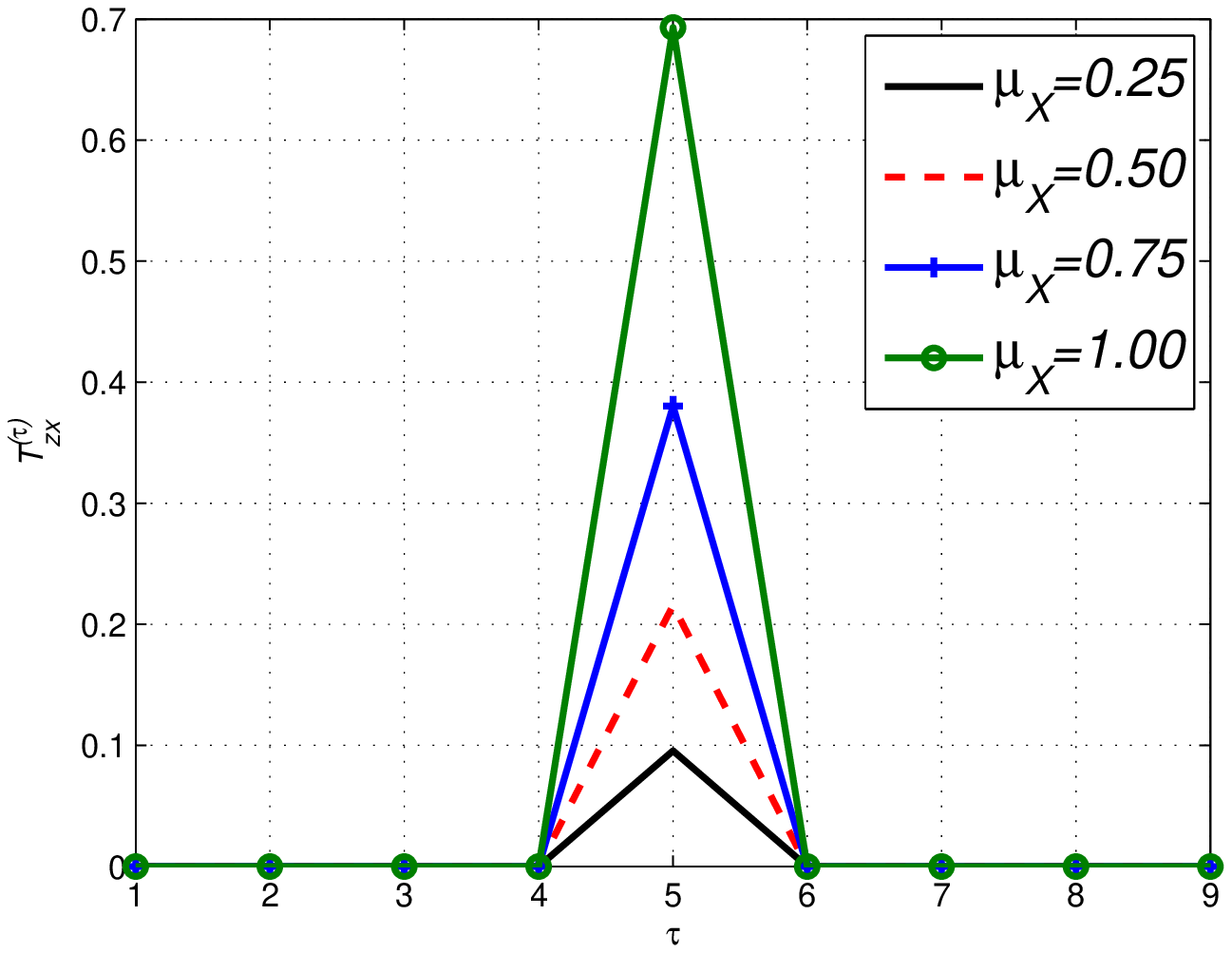}
\caption{Analytical \TE\ $T_{Z X}^{(\tau)}$ versus time lags $\tau$ of the Random Transition model with $n_s=2$ (hence $\Omega=\frac{1}{2}$) and $t_Z=5$ in equation (\ref{TzxGENERAL}) where  $\mu_X$ is varied but $\mu_Z=\frac{1}{2}$ fixed. $T_{Z X}^{(t_Z)}$ is monotonically increasing with respect to $\mu_X$. $T_{Z X}^{(t_Z)}$ is affected by $\mu_X$.}
\label{svarymx}
\end{minipage}
\hspace{0.3cm}
\begin{minipage}[b]{0.5\linewidth}
\centering
\includegraphics[width=\textwidth]{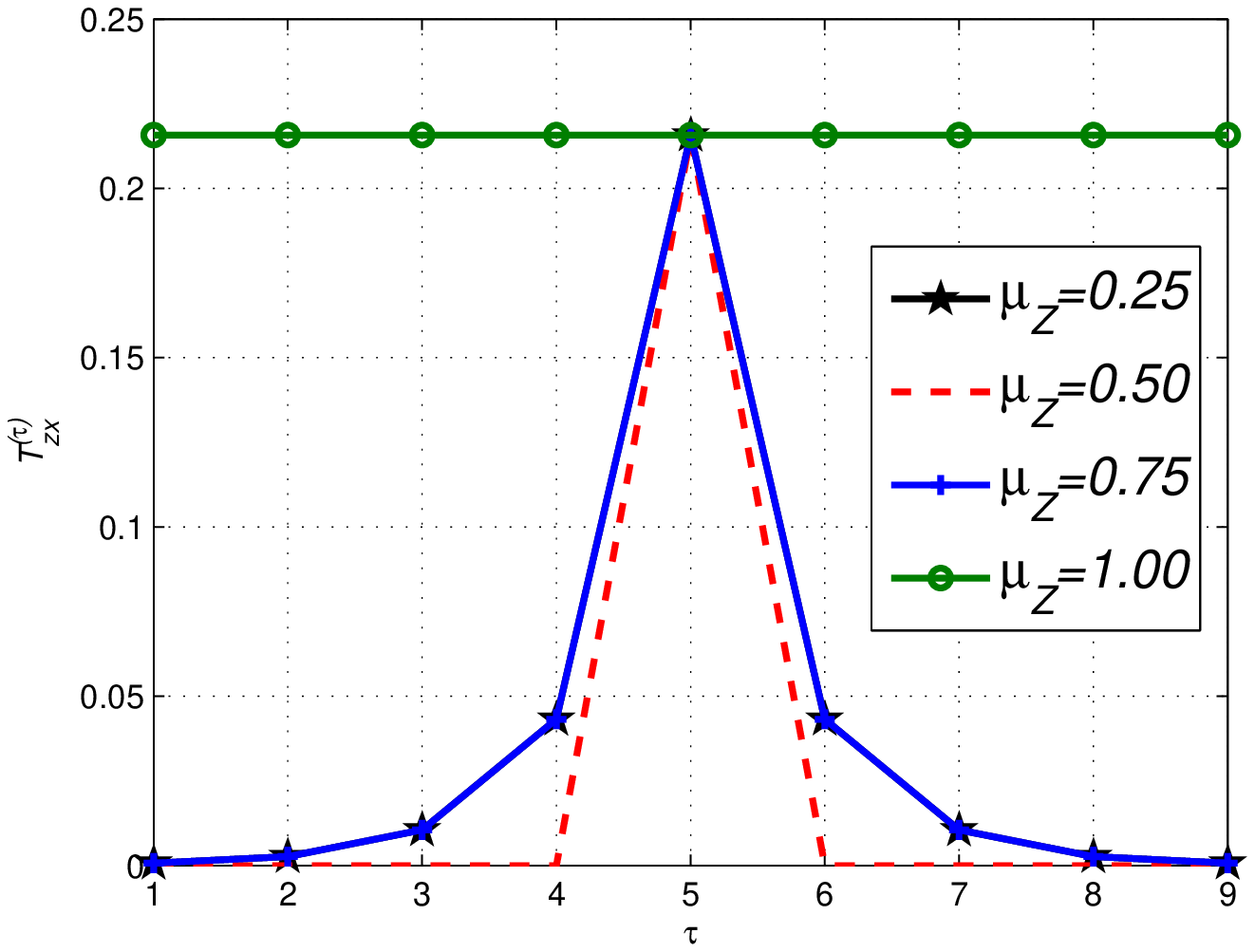}
\caption{Analytical \TE\ $T_{Z X}^{(\tau)}$ versus time lags $\tau$ of the Random Transition model with $n_s=2$ (hence $\Omega=\frac{1}{2}$) and $t_Z=5$ in equation (\ref{TzxGENERAL}) where $\mu_X=\frac{1}{2}$ fixed and $\mu_Z$ is varied.  $\mu_Z$ does not effect $T_{Z X}^{(t_Z)}$ but causes $T_{Z X}^{(\tau)} \ne 0$ at $\tau \ne t_Z$. $\mu_Z=0.25$ and $\mu_Z=0.75$ coincides}
\label{svarymz}
\end{minipage}
\end{figure}

When $n_s=2$, the model essentially replicates the \IM\ without the collective behaviour effect i.e. far above the $T_c$ where the Boltzmann distribution approaches a uniform distribution. Consequently, at these temperatures the influence of collective behaviour is close to none. One can see in Figure (\ref{svarymx}) and Figure (\ref{svarymz}) that the $\mu$ (hence the ERC) values are indeed key in determining the strength  of \TE.  In  Figure (\ref{svarymx}),  $\mu_X$ influences $T_{Z X}^{(t_Z)}$ monotonically when every other value is fixed, therefore in this case the \TE\ reflects the internal dynamics  $\mu_X$ rather than the external influence $\Omega$. If `causality' is the aim, surely $\Omega$ is the very thing that makes the relationship `causal' and should be the main focus. This is a factor that needs to be taken into account when comparing the magnitudes of \TE. Figure (\ref{svarymx}) also shows that when $\mu_Z$ is uniform (since $n_s=2$ hence $\mu_Z=\frac{1}{n_s}=\frac{1}{2}$, one gets that $T_{Z X}^{(\tau)} \ne 0$ only if $\tau=t_Z$ which makes causal lag detection fairly straight forward. However, in  Figure (\ref{svarymz}) the effect of varying $\mu_Z$ can be clearly seen in the nonzero values $T_{Z X}^{(\tau)} \ne 0$ when $\tau \ne t_Z$. Nevertheless, the value at  $\tau=t_Z$ seems to be fully determined by $\mu_X$ regardless of $\mu_Z$ value. The mechanism in which $\mu_Z$ effects $T_{Z X}^{(\tau)}$ is sketched in the appendix. 

Therefore one can conclude that when $Z$ is the source (`causal' variable) and $X$ is the target (the variable being affected by the `causal' link), the value of the \TE\ $T_{Z X}^{(\tau)}$ at $\tau=t_Z$ is influenced only by $\mu_X$ but for $\tau \ne t_Z$, $T_{Z X}^{(\tau)}$ is determined by both $\mu_X$ and $\mu_Z$. We have verified that this is indeed the case even when $n_s>2$ in this model. This should apply to all variables in the model and much more generally to any kind of source-target `causal' relationship in this sense. We suspect that this also extends to cases when there is more than one source and this will be a subject of future research. Thus for causal lag detection purposes, it is clear that theoretically \TE\ will attain maximum value at the exact causal lag. It is also clear that \TE\ at nearby lags can be nonzero due to this single `causal' relationship and on data sets it is strongly recommended to test for relative lag values. 

\subsection*{\TE\ estimations of the Random Transition model}

\begin{figure}[ht]
\begin{minipage}[b]{0.345\linewidth}
\centering
\includegraphics[width=\textwidth]{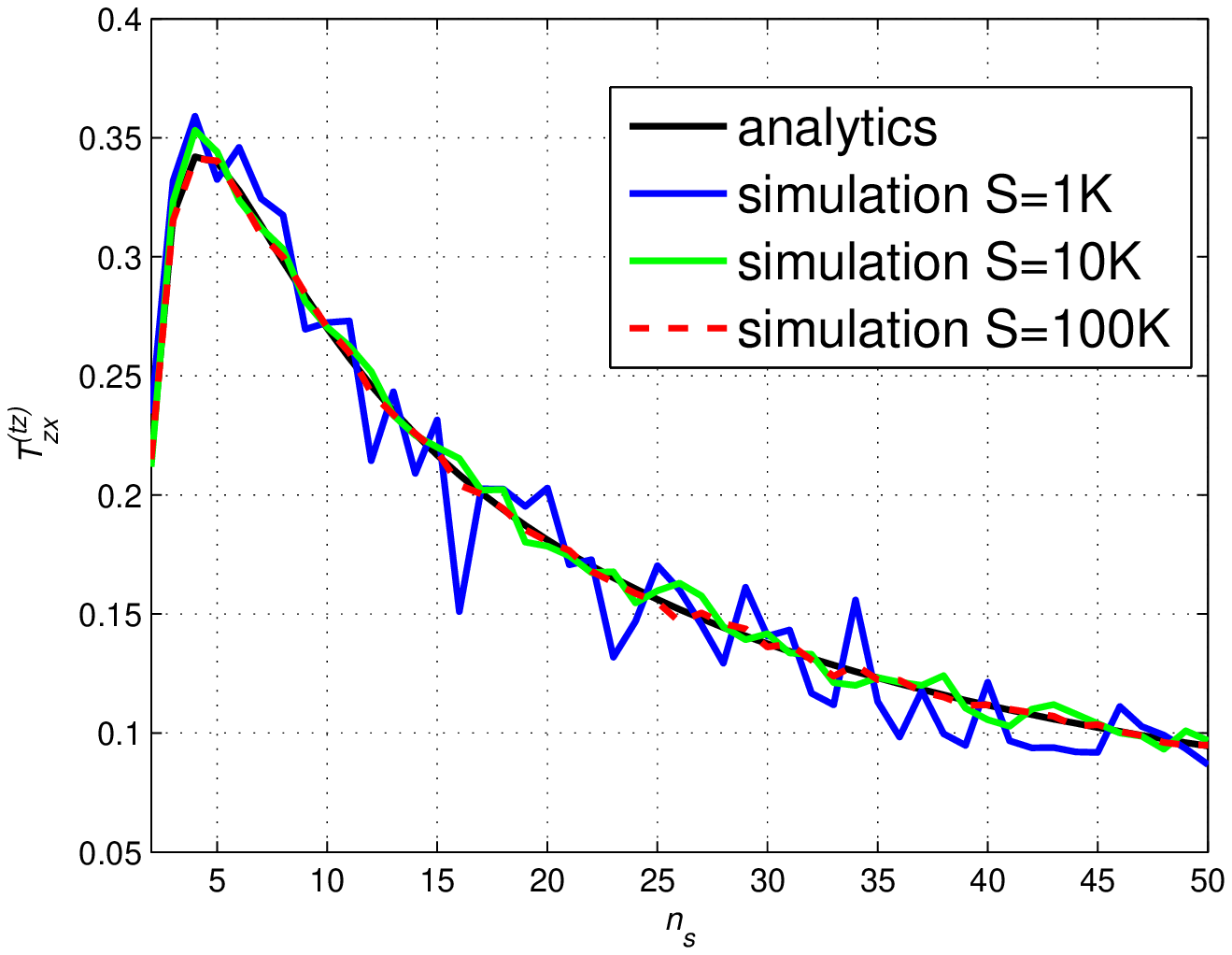}
\caption*{(a) Case $1$}
\label{case1} 
\end{minipage}
\begin{minipage}[b]{0.345\linewidth}
\centering
\includegraphics[width=\textwidth]{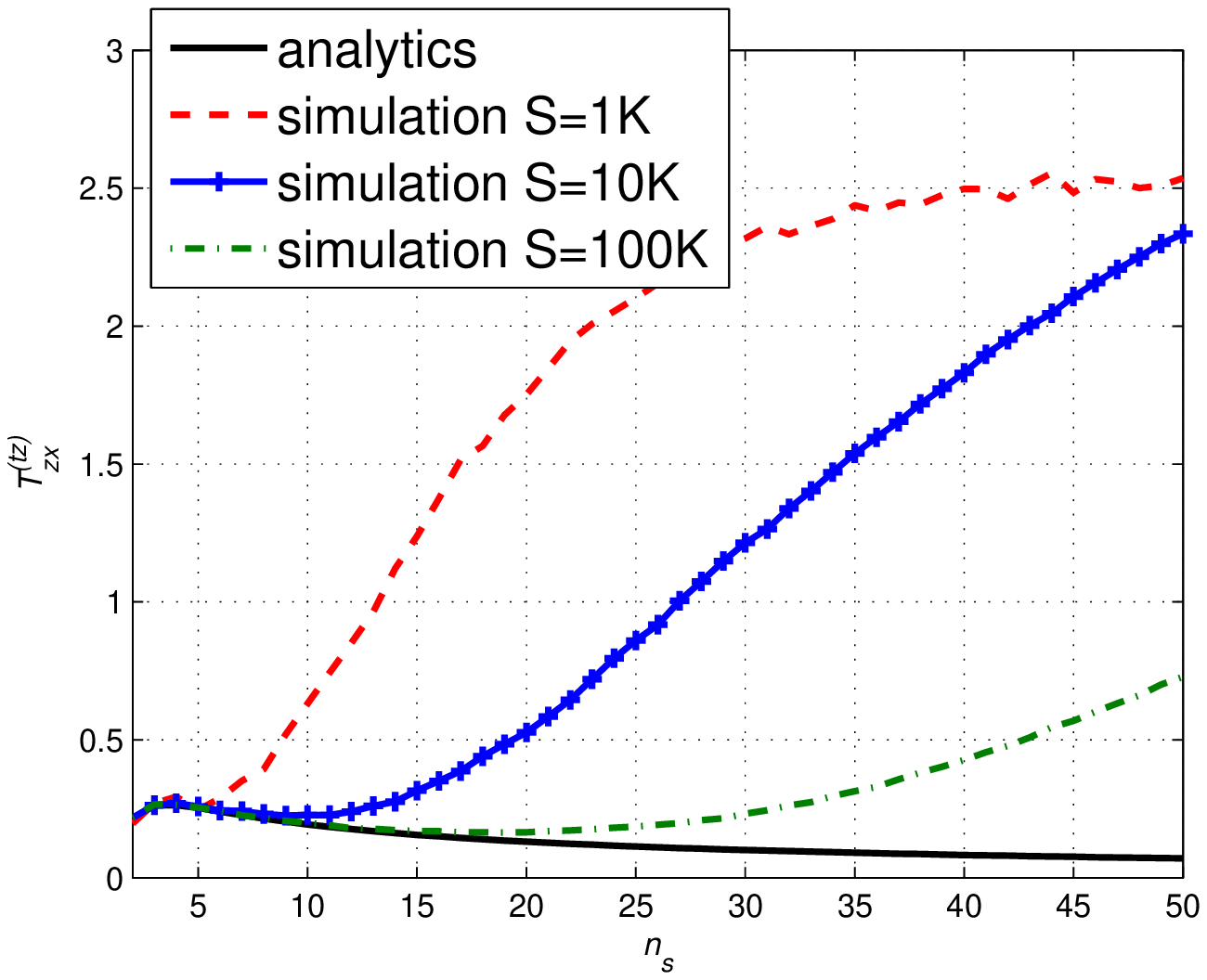}
\caption*{(b) Case $2$}
\label{case2} 
\end{minipage}
\begin{minipage}[b]{0.345\linewidth}
\centering
\includegraphics[width=\textwidth]{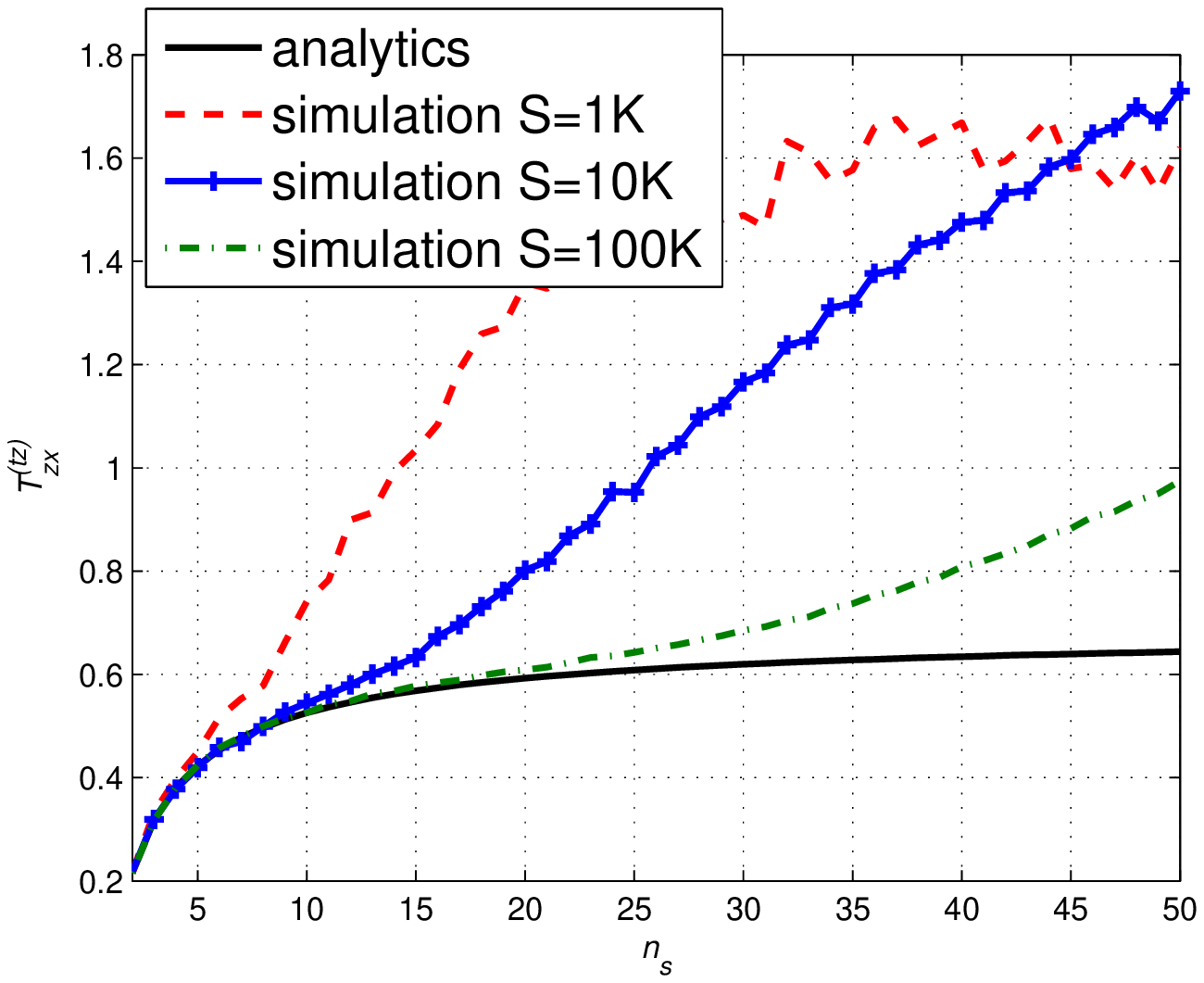}
\caption*{(c) Case $3$}
\label{case3} 
\end{minipage}
\caption{\label{cases}
\TE\ $T_{Z X}^{(t_{Z})}$ versus number of state $n_s$ for Cases $1,2$ and $3$. $\mu_X=\mu_Z=\frac{n_s-1}{n_s}$ are uniformly distributed. Analytical values obtained from substituting respective $\Omega$ values in equation (\ref{TZgeneral}). Simulated values are acquired using equation (\ref{TEtau}) on simulated data of varying sample size $S$ where $1K=1000$.}
\end{figure}

\begin{figure}[ht]
\begin{minipage}[b]{0.5\linewidth}
\centering
\includegraphics[width=\textwidth]{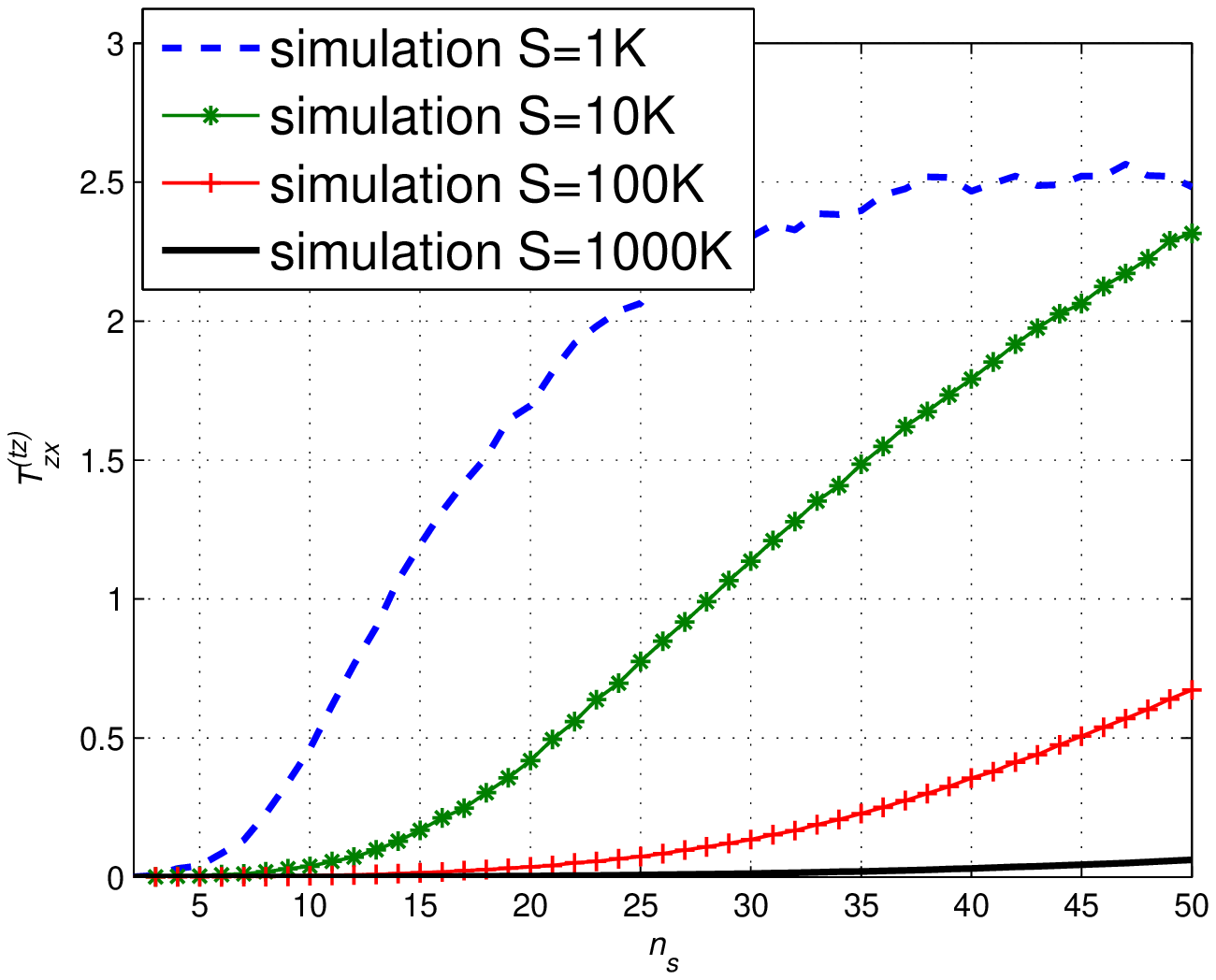}
\caption*{(a) $T_{Z X}^{(t_{Z})}$ versus $n_s$}
\label{null}
\end{minipage}
\hspace{0.5cm}
\begin{minipage}[b]{0.5\linewidth}
\centering
\includegraphics[width=\textwidth]{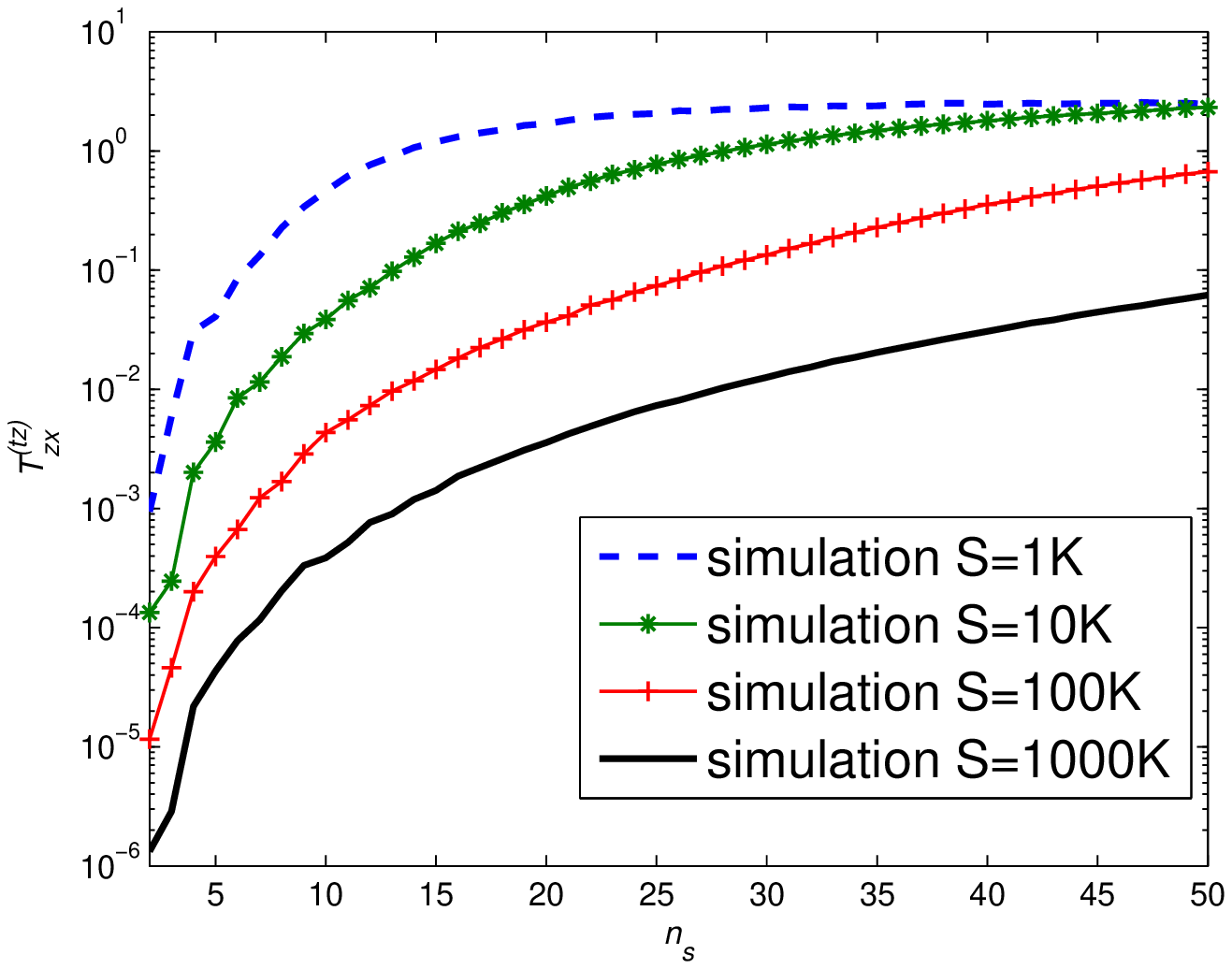}
\caption*{(b) log $T_{Z X}^{(t_{Z})}$ versus $n_s$}
\label{lognull}
\end{minipage}
\caption{\label{nullALL}
\TE\ using equation (\ref{TZgeneral}) on simulated null model with varying sample size $S$ where $1K=1000$. Analytical values are all $0$.}
\end{figure}

The estimations of \TE\ for large number of states $n_s$ requires sufficient sample size. To illustrate this we set the value $\Omega$ to three different values; $\Omega=\frac{1}{n_s}$ for Case $1$, $\Omega=\frac{n_s-1}{n_s}$ for Case $2$ and $\Omega=\frac{1}{2}$ for Case $3$. We plot the analytical \TE\,$T_{Z X}^{(t_{Z})}$, and the estimations of it on simulated values for all three cases in Figure (\ref{cases}). Even though $n_s$ is known and incorporated in the estimations, the inaccuracies are quite worrying. This situation would be even more exaggerated in situations where  $n_s$ is not known (unfortunately, this is more often than not the case). We strongly advice checking the accuracy of \TE\ estimation and adjusting the $n_s$ value  before using it for any type of analysis and drawing any conclusion. One way to do this is by generating a null model (in the case of the Random Transition model this is simply three randomly generated time series) and test the values of \TE\ as in Figure (\ref{nullALL}) to ascertain the level of accuracy that is to be expected. 

Subtracting the null model from the values on the Random Transition model is equal  to subtracting the \TE\ values of both directions as one direction is theoretically zero. However this does not quite solve the problem as the values may still be negative if the sample size is small.  There are many other types of corrections \cite{vicente11, runge12b} proposed to address this issue involving substraction of the null model in some various forms. Nevertheless, as we have seen in Figure
(\ref{dtL25}) of the amended \IM, only by subtracting the two directions of \TE\ did we obtain the clear direction as this cancelled out the underlying collective behaviour. We suspect that this will work as well for cancelling out other types of background effects and succeed in revealing directionality.



\section*{Discussion}

This paper highlights the question of distinguishing interdependencies induced by collective behaviour and individual (coupled) interactions, in order to understand the inner workings of \cs\ derived from data sets. These data sets are usually  in the form of time series that seem to behave essentially as stochastic series. It is hence of great interest to understand measures proposed to be able to probe \cau\ in view of \cs. \TE\ has been suggested as a good probe on the basis of its nonlinearities, exploratory approach and information transfer related interpretation.

To investigate the behaviour of \TE, we studied two simplistic models.
From results of applying \TE\ on the \IM, we proposed that the collective behaviour is also a type of \cau\ in the Wiener-Granger framework but highlighted that it should be identified differently from individual interactions by illustrating this issue on an amended \IM. The collective behaviour that emerges near criticality may overshadow the intrinsic directionality in the system as it is not detected by measures such as covariance (correlation) and \MI. We showed that by taking into account both directions of \TE\ on the amended \IM,  a clear direction can be identified.
In addition to that, we verified that the \TE\ is indeed maximum at the exact causal lag by utilizing the amended \IM.  

By obtaining the phase transition-like {\it difference} measure, we have shown that the \TE\ is highly dependent on the effective rate of change (ERC)
and therefore likely to be dependent on the overall activity level given by, say, the temperature in thermal systems as we demonstrated in the amended \IM. Using the Random Transition model we have illustrated that the ERC is essentially comprised of internal as well as external influences and this is why     \TE\ depicts both. This also explains why collective behaviour on the Ising model is detected as type of \cau. In complex systems where there is bound to be various interactions on top of the emergent collective behaviour, the situation can be difficult to disentangle and  caution is needed. Moreover we pointed out the danger of spurious values in the estimation of the \TE\ due to finite statistics which can be circumvented to a certain extend by a comparison of the amplitude of the causality measure in  both directions and also by use of  null models.

We believe that identifying these influences is
important for our understanding of \TE\ with the aim of utilising its
full potential in uncovering the dynamics of \cs. The mechanism of replicating \cau\ in the amended \IM\ and the Random Transition model may be used to investigate these \cau\ measures even further. Plans for future investigations involve indirect \cau, multiple sources and multiple targets. It would also be interesting to understand these measures in terms of local and global dynamics in dynamical systems. It is our hope that these investigations will help establish these \cau\ measures as a  repertoire of measures for \cs.

\section*{Appendix}

 The transition probability of the Random Transition model is as follows. We assume that if a process chooses to change it must choose one of the other states equally, thus we have that $P(X _{2}=\alpha|X_{1}=\beta, \alpha\neq\beta)=\frac{1}{n_{s}-1}P(X_2\neq X_{1})$, so that the marginal and joint probabilities remain uniform but the transition probabilities are
$$
P(X_n=\alpha|X_{n-1}=\beta)= 
\begin{cases}
1-\mu_X\Omega\ & \text{if } \alpha = \beta \\
\frac{1}{n_{s}-1}\mu_X\Omega & \text{if }\alpha \ne \beta
\end{cases}
$$
$$
P(Y_n=\alpha|Y_{n-1}=\beta) =
\begin{cases}
1-\mu_Y\Omega & \text{if } \alpha = \beta \\
\frac{1}{n_{s}-1}\mu_Y\Omega & \text{if }\alpha \ne \beta.
\end{cases}
$$
and
$$
P(Z_n=\alpha|Z_{n-1}=\beta)=
\begin{cases}
1-\mu_Z & \text{if } \alpha = \beta \\
\frac{1}{n_{s}-1}\mu_Z & \text{if }\alpha \ne \beta
\end{cases}
$$
where $\Omega=P( \text{ condition fulfilled })$ such that one can control `dependence' on $Z$  by altering $\Omega$.

\subsection*{The relationship between $\Omega$ and $Q$}\label{omegaNq}

To understand how the values of $\mu_Z$ affects the value of $T^{(\tau)}_{ZX}$ we need a different variable. Let $Q$ be the probability that the condition is fulfilled given current knowledge at time $\tau$ such that 
$Q_{sgn(\gamma)}^{(\tau)}=P(\text{ condition fulfilled  }| \text{ knowledge at time } \tau)$. The value of $Q_{sgn(\gamma)}^{(\tau)}$  will depend on $\gamma$, and in our model here, particularly on whether or not $Z_{n-t_z}=\gamma$ satisfies the condition. One can  divide the possible states $\gamma$ of all the processes into two groups such that 
$$G_U=\{\gamma \in A, Z_{n-t_Z}=\gamma \text{ fulfills the condition}\} \quad \text{and}$$ 
$$G_D=\{\gamma \in A, Z_{n-t_Z}=\gamma \text{ does not fulfill the condition} \}.$$
Note that $|G_U|=n_{s}\Omega$ and $|G_D|=n_{s}(1-\Omega)$  since $\Omega=P(\text{ condition fulfilled })$ such that $\Omega$ can be interpreted as the proportion of states of $Z$ that fulfill the condition. Due to equiprobability of spins and uniform initial distribution, for any $\tau$ there are only two possible values of $Q_{sgn(\gamma)}^{(\tau)}$, one for $\gamma \in G_U$ and one for $\gamma \in G_D$. Therefore define $sgn(\gamma)$ such that
\begin{align}\label{sgn}
sgn(\gamma)=
\begin{cases}
        + & \text{if } \gamma \in G_U \\
        - & \text{if } \gamma \in G_D
\end{cases}
\end{align}
to get
\begin{align}\label{Qgeneral}
Q_{sgn(\gamma)}^{(\tau)}=
\begin{cases}
        Q_{+}^{(\tau)} & \text{if } \gamma \in G_U \\
        Q_{-}^{(\tau)} & \text{if } \gamma \in G_D.
\end{cases}
\end{align}
 Thus 
$Q_{sgn(\gamma)}^{(\tau)}=P(\text{ condition fulfilled  }|Z_{n-\tau}=\gamma)$
with the $sgn(\gamma)$ as in equation (\ref{sgn}).

The relationship between $Q_{sgn(\gamma)}^{(\tau)}$ and $\Omega$ can be defined using the formula for total probability 
$
P(B)=\sum_\gamma P(B|Z=\gamma)P(Z=\gamma).
$
Let $B=\{\text{ condition fulfilled }\}$ and using the fact that $P(Z_{n-\tau}=\gamma)=\frac{1}{n_s}$, we get that
\begin{align}\label{omegaNQ}
\Omega=P(B)
=\sum_\gamma P(B|Z_{n-\tau}=\gamma)P(Z_{n-\tau}=\gamma)
=\frac{1}{n_{s}}\sum_\gamma Q_{sgn(\gamma)}^{(\tau)}.
\end{align}
Due to the sole dependence of $Z$ on $\mu_Z$, $\mu_Z=\frac{n_s-1}{n_s}$ will make the transition probability of $Z$ uniform such that $P(Z_n=\alpha|Z_{n-1}=\beta)=\frac{1}{n_{s}}$ for any $n$ since
we have that$$
P(Z_n=\alpha|Z_{n-1}=\beta)=
\begin{cases}
1-\mu_Z=1-\frac{n_s-1}{n_s}=\frac{1}{n_{s}} & \text{if } \alpha = \beta \\
\frac{1}{n_{s}-1}\mu_Z=\frac{1}{n_{s}-1}\frac{n_s-1}{n_s}=\frac{1}{n_{s}} & \text{if }\alpha \ne \beta
\end{cases}
$$
for any $\alpha, \beta \in A=\{1, \cdots, n_s \}$. Consequently, $\mu_Z=\frac{n_s-1}{n_s}$ also makes all values of $Q_{sgn(\gamma)}^{(\tau)}$ uniform so that equation (\ref{omegaNQ}) becomes
\begin{align}\label{omegaequalq}
\Omega=\frac{1}{n_{s}}\sum_\gamma Q_{sgn(\gamma)}^{(\tau)}=\frac{1}{n_{s}} n_{s}Q_{sgn(\gamma)}^{(\tau)}=Q_{sgn(\gamma)}^{(\tau)}.
\end{align}
Therefore on the model when the $\mu_Z=\frac{n_s-1}{n_s}$, we have that $\Omega=Q^{(\tau)}_{sgn(\gamma)}$ for any $\tau=t_Z$.  And this is why we get Figure (\ref{svarymx}), where  $T_{Z X}^{(\tau)} \ne 0$ only if $\tau=t_Z$ since $\Omega=Q^{(\tau)}_{sgn(\gamma)}$ in equation (\ref{TzxGENERAL}) cancels out. 

For any $\mu_Z$, the relationship between $Q_{+}^{(\tau)}$ and $Q_{-}^{(\tau)}$ can be derived from  equation (\ref{omegaNQ})
where
\begin{align}
n_s\Omega=\sum_\gamma Q_{sgn(\gamma)}^{(\tau)}
&=\sum_{\gamma \in G_U} Q_{sgn(\gamma)}^{(\tau)}+\sum_{\gamma \in G_D} Q_{sgn(\gamma)}^{(\tau)}
=|G_U|Q_+^{(\tau)}+|G_D|Q_-^{(\tau)}\\\notag
n_s\Omega&= n_s\Omega\,\ Q_+^{(\tau)}+ n_s(1-\Omega)Q_-^{(\tau)}\\\notag
\Omega(1-Q_+^{(\tau)})&=(1-\Omega)Q_-^{(\tau)}
\end{align}
Note that when $n_s=2$ (hence $\Omega=\frac{1}{2}$) this simplifies to $ Q_{+}^{(\tau)}+ Q_{-}^{(\tau)}=1$. 

\subsection*{\TE\ formula on the Random Transition model}

Using 
$Q_{sgn(\gamma)}^{(\tau)}$ as in equation (\ref{Qgeneral}) we have that $$
\frac{P(X_n=\alpha|X_{n-1}=\beta,Z_{n-\tau}=\gamma)}{P(X_n=\alpha|X_{n-\tau}=\beta)} =
\begin{cases}
\frac{1-\mu_XQ_{sgn(\gamma)}^{(\tau)}}{1-\mu_X\Omega} & \text{if } \alpha = \beta \\
\frac{\frac{1}{n_{s}-1}\mu_XQ_{sgn(\gamma)}^{(\tau)}}{\frac{1}{n_{s}-1}\mu_X\Omega}=\frac{Q_{sgn(\gamma)}^{(\tau)}}{\Omega} & \text{if }\alpha \ne \beta,
\end{cases}
$$
which gives us
\begin{align} \label{TzxGENERAL}
T_{Z X}^{(\tau)}=&\sum _{\alpha}\sum _{\beta}\sum _{\gamma}P(X_{n}=\alpha, X_{n-1}=\beta,Z_{n-\tau}
=\gamma)log \, \frac{P(X_{n}=\alpha|X_{n-1}=\beta,Z_{n-\tau}=\gamma)}{P(X_{n}=\alpha|X_{n-1}=\beta)}\notag\\
=& |\{X_n=X_{n-1}\}|\sum_{\gamma}\left[\frac{1-\mu_XQ_{sgn(\gamma)}^{(\tau)}}{n^{2}_{s^{}}}\,log\,\frac{1-\mu_XQ_{sgn(\gamma)}^{(\tau)}}{1-\mu_X\Omega}
\right] \notag\\
&+|\{X_n \ne X_{n-1}\}|\sum_{\gamma}\left[\frac{\frac{1}{n_{s}-1}\mu_XQ_{sgn(\gamma)}^{(\tau)}}{n^{2}_{s^{}}}\,log\,\frac{Q_{sgn(\gamma)}^{(\tau)}}{\Omega}\right] \notag\\
=& n_s\sum_{\gamma}\left[\frac{1-\mu_XQ_{sgn(\gamma)}^{(\tau)}}{n^{2}_{s^{}}}\,log\,\frac{1-\mu_XQ_{sgn(\gamma)}^{(\tau)}}{1-\mu_X\Omega}
\right] \notag\\
&+n_s(n_s-1)\sum_{\gamma}\left[\frac{\frac{1}{n_{s}-1}\mu_XQ_{sgn(\gamma)}^{(\tau)}}{n^{2}_{s^{}}}\,log\,\frac{Q_{sgn(\gamma)}^{(\tau)}}{\Omega}\right] \notag\\
=& \frac{1}{n_s}\sum_{\gamma \in G_U }\left[(1-\mu_X Q_{sgn(\gamma)}^{(\tau)})\,log\,\frac{1-\mu_XQ_{sgn(\gamma)}^{(\tau)}}{1-\mu_X\Omega}
+\mu_X Q_{sgn(\gamma)}^{(\tau)}\,log\,\frac{Q_{sgn(\gamma)}^{(\tau)}}{\Omega}\right]\notag\\
&+\frac{1}{n_s}\sum_{\gamma \in G_D}\left[(1-\mu_X Q_{sgn(\gamma)}^{(\tau)})\,log\,\frac{1-\mu_XQ_{sgn(\gamma)}^{(\tau)}}{1-\mu_X\Omega}
+\mu_X Q_{sgn(\gamma)}^{(\tau)}\,log\,\frac{Q_{sgn(\gamma)}^{(\tau)}}{\Omega}\right] \notag\\
=& \frac{1}{n_s}(n_s \Omega)\left[(1-\mu_X Q_{+}^{(\tau)})\,log\,\frac{1-\mu_XQ_{+}^{(\tau)}}{1-\mu_X\Omega}
+\mu_X Q_{+}^{(\tau)}\,log\,\frac{Q_{+}^{(\tau)}}{\Omega}\right]\notag\\
&+\frac{1}{n_s}n_s(1-\Omega)\left[(1-\mu_X Q_{-}^{(\tau)})\,log\,\frac{1-\mu_XQ_{-}^{(\tau)}}{1-\mu_X\Omega}
+\mu_X Q_{-}^{(\tau)}\,log\,\frac{Q_{-}^{(\tau)}}{\Omega}\right] \notag\\
=&\, \Omega\left[(1-\mu_X Q_{+}^{(\tau)})\,log\,\frac{1-\mu_XQ_{+}^{(\tau)}}{1-\mu_X\Omega}
+\mu_X Q_{+}^{(\tau)}\,log\,\frac{Q_{+}^{(\tau)}}{\Omega}\right]\notag\\
&+(1-\Omega)\left[(1-\mu_X Q_{-}^{(\tau)})\,log\,\frac{1-\mu_XQ_{-}^{(\tau)}}{1-\mu_X\Omega}
+\mu_X Q_{-}^{(\tau)}\,log\,\frac{Q_{-}^{(\tau)}}{\Omega}\right]
\end{align}
where we used
the  Bayes theorem i.e
\begin{align*}
P(X_{n}=\alpha,Z_{n-1}=\gamma,X_{n-1}=\beta)=\frac{1}{n_{s}^2}P(X_{n}=\alpha|Z_{n-1}=\gamma,X_{n-1}=\beta).
\end{align*}
Due to independence, if $Y$ were to be conditioned on $X$ we would have that
$$
\frac{P(Y_n=\alpha|Y_{n-1}=\beta,X_{n-\tau}=\gamma)}{P(Y_n=\alpha|Y_{n-1}=\beta)} =\frac{P(Y_n=\alpha|Y_{n-1}=\beta)}{P(Y_n=\alpha|Y_{n-1}=\beta)} =1.
$$
Therefore for values other than when $X$ and $Y$  conditioned on $Z$, this ratio will yield $1$. This renders $T_{X Z}^{(\tau)}=T_{Y Z}^{(\tau)}=T_{Y X}^{(\tau)}=T_{X Y}^{(\tau)}=0$. And if we get that $T_{Z X}^{(\tau)} \ne 0$, we can say that Transfer Entropy indicates \cau\ or some form of directionality from $Z$ to $X$ and $Z$ to $Y$, at time lag $\tau$. In a similar manner for $\alpha,\beta,\gamma \in A$ we have that
$$
\frac{P(Y_n=\alpha|Y_{n-1}=\beta,Z_{n-\tau}=\gamma)}{P(Y_n=\alpha|Y_{n-1}=\beta)} =
\begin{cases}
\frac{1-\mu_YQ_{sgn(\gamma)}^{(\tau)}}{1-\mu_Y\Omega} & \text{if } \alpha = \beta \\
\frac{\frac{1}{n_{s}-1}\mu_YQ_{sgn(\gamma)}^{(\tau)}}{\frac{1}{n_{s}-1}\mu_Y\Omega}=\frac{Q_{sgn(\gamma)}^{(\tau)}}{\Omega} & \text{if }\alpha \ne \beta
\end{cases}
$$
such that $T_{ZY}^{(\tau)}$ in exactly like equation (\ref{TzxGENERAL}) except that  $\mu_X$ is replaced with $\mu_Y$.

When $\tau=t_Z$ we have that $Q_{sgn(\gamma)}^{(t_{Z})}$ is either $0$ or $1$ since the condition was placed at $n-t_Z$. More specifically we will have that  $Q_{+}^{(t_{Z})}=1$ and that $Q_{-}^{(t_{Z})}=0$. Putting these two values in equation (\ref{TzxGENERAL}) we obtain
\begin{align}\label{TZgeneral}
T_{Z X}^{(t_{Z})}=&\, \Omega\left[(1-\mu_X Q_{+}^{(t_{Z})})\,log\,\frac{1-\mu_XQ_{+}^{(t_{Z})}}{1-\mu_X\Omega}
+\mu_X Q_{+}^{(t_{Z})}\,log\,\frac{Q_{+}^{(t_{Z})}}{\Omega}\right]\notag\\
&+(1-\Omega)\left[(1-\mu_X Q_{-}^{(t_{Z})})\,log\,\frac{1-\mu_XQ_{-}^{(t_{Z})}}{1-\mu_X\Omega}
+\mu_X Q_{-}^{(t_{Z})}\,log\,\frac{Q_{-}^{(t_{Z})}}{\Omega}\right]\notag\\
=\,\Omega&(1-\mu_X)\,log\,\frac{1-\mu_X}{1-\mu_X\Omega}+\,\Omega\mu_X\,log\,\frac{1}{\Omega}
+(1-\Omega)\,log\,\frac{1}{1-\mu_X\Omega}.
\end{align}
A more thorough treatment of the Random Transition model and other methods of \TE\ estimations is given in \cite{fatimah}.
\section*{Acknowledgments}


\section*{Figure Legends}

\begin{figure}[ht]
\begin{minipage}[b]{0.5\linewidth}
\centering
\includegraphics[width=\textwidth]{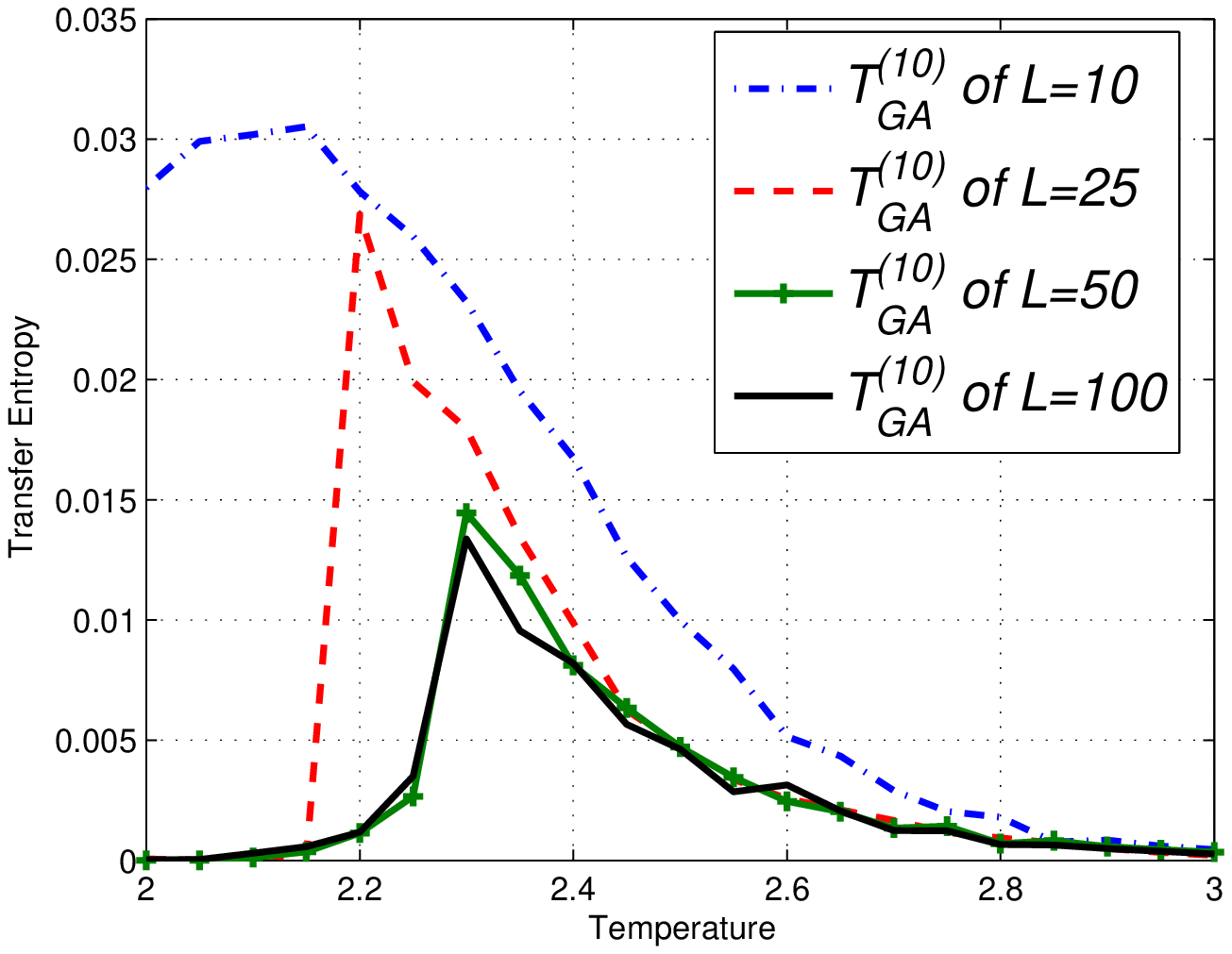}
\caption{\textbf{\TE\ $T^{(10)}_{GA}$ on the \IM} of lengths $L=10,25,50,100$ obtained using equation (\ref{TEtau}). Peaks can be seen at respective $T_c$.}
\label{TEGA100}
\end{minipage}
\hspace{0.3cm}
\begin{minipage}[b]{0.5\linewidth}
\centering
\includegraphics[width=\textwidth]{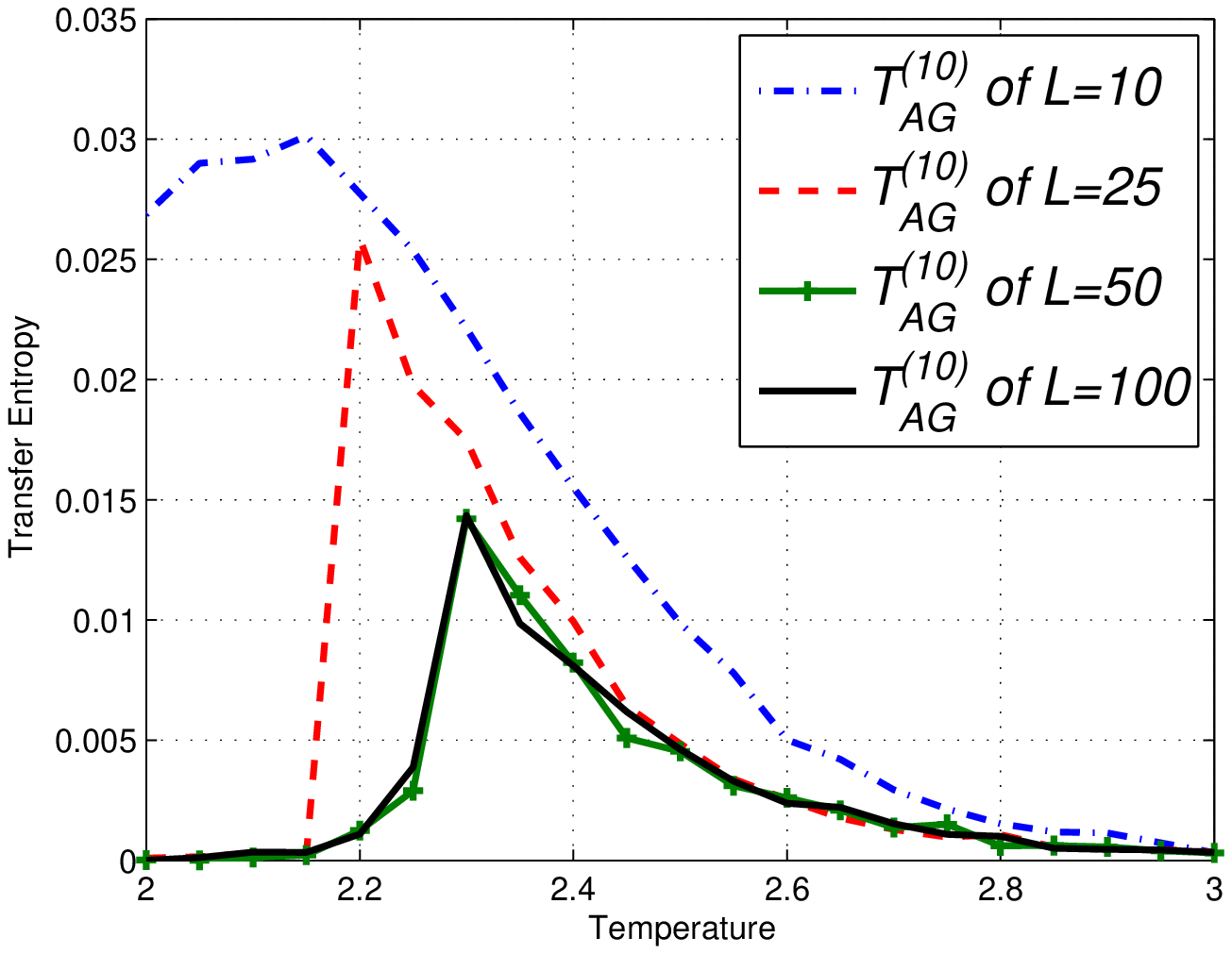}
\caption{\textbf{\TE\ $T^{(10)}_{AG}$ on the \IM} of lengths $L=10,25,50,100$ obtained using equation (\ref{TEtau}). Peaks can be seen at respective $T_c$.}
\label{TEAG100}
\end{minipage}
\end{figure}


\end{document}